\documentclass[12pt,preprint]{aastex} 
\shorttitle{} 
\shortauthors{} 
 
\begin{document} 
 
\received{} 
\accepted{} 
 
\title{KIC 4142768: An Evolved Gamma Doradus/Delta Scuti Hybrid Pulsating Eclipsing Binary with Tidally Excited Oscillations}  
 
\author{Zhao Guo$^{1,2}$, Jim Fuller$^{3}$, Avi Shporer$^{4}$, Gang Li$^{5}$, Kelly Hambleton$^{6}$, Joseph Manuel$^{6}$, Simon Murphy$^{5}$, Howard Isaacson$^{7}$} 
\affil{${1}$ Center for Exoplanets and Habitable Worlds, Department of Astronomy \& Astrophysics, 525 Davey Laboratory, The Pennsylvania State University, University Park, PA 16802, USA \\
${2}$ Copernicus Astronomical Center, Polish Academy of Sciences, Bartycka 18, 00-716 Warsaw, Poland \\
${3}$ TAPIR, Mailcode 350-17, California Institute of Technology, Pasadena, CA 91125, USA \\
${4}$ Department of Physics and Kavli Institute for Astrophysics and Space Research, Massachusetts Institute of Technology, Cambridge, MA 02139, USA \\
${5}$ Sydney Institute for Astronomy (SIfA), School of Physics, The University of Sydney, NSW 2006, Australia \\
${6}$ Department of Astrophysics and Planetary Science, Villanova University, 800 East Lancaster Avenue, Villanova, PA 19085, USA \\
${7}$ Department of Astronomy, University of California, Berkeley CA 94720, USA \\
}

\slugcomment{09/03/2019} 

 
\begin{abstract} 
We present the characterization of KIC 4142768, an eclipsing binary with two evolved A-type stars in an eccentric orbit with a period of 14 days. We measure the fundamental parameters of the two components ($M_1=2.05M_{\odot}, R_1=2.96R_{\odot}$ and $M_2=2.05M_{\odot}, R_2=2.51R_{\odot}$) by combining {\it Kepler} photometry and spectra from {\it Keck} HIRES. The measured surface rotation rates are only one-fifth of the pseudo-synchronous rate of the eccentric orbit. Fourier spectrum of the light curve reveals hybrid pulsations of $\delta$ Scuti and $\gamma$ Doradus type, with pulsation frequencies at about $15-18$ day$^{-1}$ for p modes and about $0.2-1.2$ day$^{-1}$ for low-frequency g modes. Some of the g modes are exact orbital harmonics and are likely tidally excited.  Their pulsation amplitudes and phases both agree with predictions from the linear tidal theory for $l=2, m=2$ prograde modes. We examine the period spacing patterns in the free oscillating g modes and identify them mostly as prograde sectoral dipole modes. The unstable frequency range and frequency spacing of p modes and the inferred asymptotic g-mode period spacings both agree with the stellar model for the primary star evolved to a late stage of the main sequence. The inferred rotation rate of the convective core boundary is very slow, similar to the small surface rotation rate inferred from the spectroscopy. The measured surface and near-core rotation rates provide constraints for testing the mechanism of angular momentum transfer and tidal synchronization in evolved eccentric binary star systems. 
\end{abstract} 
 
 
 

\section{Introduction}

$\delta$ Scuti ($\delta$ Sct) type variable stars, being relatively numerous and luminous, were identified as a class of pulsators early (Breger 1979; Rodriguez et al.\ 2000). $\gamma$ Doradus ($\gamma$ Dor) variables, pulsating in higher-order gravity modes, were identified to be another class of pulsators in the late 20th century (Breger \& Beichbuchne 1996; Kaye et al.\ 1999). A small number of stars are called `hybrids' since they show both types of pulsations (Handler \& Shobbrook 2002). Precise photometry from space lowered the detection threshold and found many $\delta$ Sct/$\gamma$ Dor hybrids (Grigahcene et al.\ 2010). Actually, Balona et al.\ (2015) found that low frequencies are present in most $\delta$ Scuti stars observed by {\it Kepler}. State-of-the-art convection theory shows that both the radiative $\kappa$ mechanism and the coupling between convection and oscillations play a major role in the excitation of $\delta$ Scuti and $\gamma$ Dor stars, with the former mainly for warmer $\delta$ Sct stars and the later for cooler $\delta$ Sct and $\gamma$ Dor stars (e.g., Xiong et al.\ 2016).

In general, the observed frequency range of $\delta$ Sct stars matches the theoretical range of unstable modes calculated from stellar models with current opacities (Pamyatnykh2003; Casas et al.\ 2009; Zwintz et al.\ 2014). For $\gamma$ Dor stars, the comparison between observation and theory is by far mostly restricted to the instability strip (Dupret et al.\ 2004; Bouabid et al.\ 2013; Xiong et al.\ 2016), not the frequency range of unstable modes of individual stars (see Maceroni 2014 for an exception). This is, again, due to the difficulty in modeling the interaction between the convection and oscillations.

A crucial step in asteroseismology is the successful identification of oscillation modes. This relies heavily on recognizing patterns in the observed oscillation frequencies. For $\delta$ Sct stars, the pulsation spectrum does not generally show regularities, although some regular patterns have been found and interpreted as the large frequency separation (Garcia-Hernandez et al.\ 2015). Oscillation calculations of 2D stellar structure models support this interpretation (Reese et al.\ 2017). $\gamma$ Dor stars are more amenable to reveal patterns since their g modes are in the asymptotic regime and nearly equally-spaced in period. Indeed, the period spacing ($\Delta P$) and period ($P$) diagram have been obtained for hundreds of $\gamma$ Dor stars and used to derive the internal rotation rates and asymptotic period spacings (Bedding 2015; Saio et al.\ 2015; Van Reeth et al.\ 2016; Ouazzani et al.\ 2017; Li et al.\ 2019). Christophe et al.\ (2018) stretched the oscillating periods of g modes so that they are equidistant, which facilitates the identification of regularities with the Fourier technique. For a nice reference on the period spacing behavior of g modes, please refer to Miglio et al.\ (2008).

If pulsation amplitudes and phases can be obtained in different photometric passbands, the sensitivity of limb darkening to different pulsating modes (spherical harmonics) can also reveal the mode identification (Balona \& Evers 1999; Garrido 2000; Dupret et al.\ 2003). Spectroscopic mode identification can also be performed, but it requires extensive and high-resolution spectroscopic observations of spectral lines. This has been applied to $\delta$ Sct stars (Mathias et al.\ 1997; Kennelly et al.\ 1998; Zima et al. 2006) and $\gamma$ Dor stars (Brunsden et al.\ 2012, 2018).
In some rare cases, rotational splittings of oscillation modes can help the mode identification, but this is restricted to slow rotators.

A step further in asteroseismology is to model individual frequencies. For $\delta$ Sct stars, seismic modeling is difficult in general. Attempts by using the perturbative method to rotation (Suarez et al.\ (2005) for Altair and Pamyatnykh (1998) for XX Pyx) or 2D models (Deupree et al.\ (2011, 2012) for $\alpha$ Oph) only have very limited success. Other difficulties such as mode selection and non-linear mode coupling prevent us from obtaining a satisifactory seismic model. In fact, the review by Balona (2010) states that `no entirely satisfactory asteroseismic solution has emerged for any $\delta$ Sct stars'\footnote{Recently, Bedding et al. show that it may be possible for some high-frequency `nice' $\delta$ Scuti stars.}. On the other hand, individual frequencies of $\gamma$ Dor stars are not usually exploited. Most of the asteroseismic modeling of $\gamma$ Dor stars only use the period spacing in the merit function (e.g., Schmid \& Aerts (2016) and Saio et al. (2015)).

It is advantageous to study pulsating stars in binaries, especially eclipsing binaries since the synergy can provide us with accurate fundamental stellar parameters and refine our knowledge on stellar physics. Chapellier et al.\ (2012) and Chapellier \& Mathias (2013) measured the fundamental stellar parameters of two $\delta$ Sct/$\gamma$ Dor hybrid binaries observed by {\it CoRoT}. Schmid et al.\ (2015), Schmid \& Aerts (2016), and Keen et al.\ (2015) studied $\delta$ Sct/$\gamma$ Dor hybrids in the binary KIC 10080943 and performed seismic modeling of  g modes. Hełminiak et al.\ (2017) presented the study of the hybrid pulsator in a hierarchical system KIC 4150611. The studies on hybrid pulsating binaries also include Maceroni et al.\ (2014), Hambleton et al.\ (2013), and  Guo et al.\ (2016, 2017a), Lampens (2018), etc. Additionally, the improvement of our tidal theory relies on the study of orbital evolution of binaries. For stars with radiative envelopes in binaries, the dominant dissipation mechanism is the radiative damping of gravity-modes excited by the dynamical tide (Zahn 1975). The effect of dynamic tides can be revealed as tidally excited oscillations in the observed flux. This has been observed in many binary stars observed by space missions such as Kepler, BRITE and TESS (Welsh et al.\ 2011; Thompson et al.\ 2012; Hambleton et al.\ 2016, 2018; Guo et al.\ 2017b; Fuller 2017; Pablo et al. 2017; Jayasinghe et al.\ 2019).

In this paper, we study an eccentric binary system which shows both $\delta$ Sct/$\gamma$ Dor type self-driven oscillations and tidally forced oscillations. We outline as follows. Section 2 presents the binary modeling by combining the {\it Kepler} photometry and ground-based spectroscopy. Section 3 concerns the evolution stage of this binary. In Section 4, we present a detailed asteroseismic interpretation of both the high-frequency p modes and low-frequency g modes including tidally excited modes. The analysis on pulsations solidifies convincingly the previous measured binary parameters.
After commenting on a previous study of this binary in Sec.\ 5, we discuss the implications of this work and future prospects in the last section.

\section{Binary Modeling with Kepler Photometry and Keck HIRES Spectroscopy}

KIC 4142768 ($\alpha$=19:09:03.08, $\delta$=+39:15:36.1) was observed by the {\it Kepler} satellite from Quarter 0 through 17 (1460 days). It was included in the Kepler Eclipsing Binary Catalog (KEBC) (Prsa et al.\ 2011; Slawson et al.\ 2011; Kirk et al.\ 2016). As noted by Balona (2018), the listed orbital period ($P=27.9916030$d) is incorrect and the true period is actually half of the listed value. In KEBC, KIC 4142768 is flagged as a heartbeat binary `HB' with tidally induced pulsations `TP'. Armstrong et al.\ (2014) derived the effective temperature by fitting the spectral energy distribution ($T_{\rm eff1}=5435\pm359K, T_{\rm eff2}=7698\pm842K$) and the Kepler-INT Survey (KIS) temperatures listed in Greiss et al.\ (2012) are $T_{\rm eff1}=6302\pm737K, T_{\rm eff2}=7017\pm1257K$. KIC 4142768 has a {\it Kepler} magnitude of $K_p=12.12$ and only long cadence data are available. We obtained the Simple Aperture Photometry (SAP) light curves from the Mikulski Archive for Space Telescopes (MAST) database and prepared the raw light curves following procedures in our previous papers (Guo et al.\ 2016; 2017a,b). We did not consider the contamination effect since the reported values in MAST are zero in all quarters. 

We obtained high-resolution spectra (R $\approx 55000$)  from the HIRES spectrograph on Keck. The instrumental setup of the California Planet Search (CPS) was used (Howard 2009). Please refer to Section 2.2 in Shporer et al.\ (2016) and Petigura et al.\ (2017) for details on the spectral reduction pipeline. We use the reduced, wavelength-calibrated product for subsequent analysis.
The spectra are double-lined and clearly reveal the binary nature of this system (Figure 1). We derived the radial velocities (RVs) of the two components by cross-correlating the observed spectra with a template generated from the Kurucz-model based BLUERED library (Bertone et al.\ 2008). The library has a fixed mixing length parameter $l/H_p = 1.25$ and the microturbulence velocity of $2$ km s$^{-1}$. The original library spectra are broadened and limb-darkened with the rotational kernel in Gray (2008). The linear limb-darkening coefficients in Claret \& Bloemen (2011) are adopted. We find the Echelle order spanning the wavelength range $5120-5220 $$\mathrm{\AA}$ can give the least scatter in the derived RVs. Using the measured radial velocities listed in Table 3, we then separate the observed composite spectra to two individual spectra by using the tomographic algorithm in Bagnuolo et al.\ (1994). The separated spectra of both components are compared to a grid of BLUERED spectra to obtain the optimized atmosphere parameters ($T_{\rm eff}$, $\log g$, [Fe/H]) and $v \sin i$. The optimization was performed by using both the genetic algorithm PIKAIA (Charbonneau 1995) and the MCMC sampler {\it emcee} (Foreman-Mackey et al.\ 2013). The above steps were iterated once to obtain an improved solution. The final results are shown in Figure 1 and listed in Table 1. We find the two stars in this binary have similar atmospheric parameters, with ($T_{\rm eff}$, $\log g$, $v \sin i$)=(7327 K, 3.53 dex, 8.7 km s$^{-1}$) for the primary star and  (7283 K, 3.51 dex, 7.0 km s$^{-1}$) for the secondary. Both components show slightly sub-solar metallicity ($[Fe/H]=-0.02$).

We then opt to find a binary star model by fitting both the Kepler light curve (LC) and the RVs with the Eclipsing Light Curve (ELC) code (Orosz \& Hilditch 2002). ELC implements the Rochel model and Phoenix atmosphere model and fully account for the tidal distortion and reflection effect. We fix the effective temperature of the primary star ($T_{\rm eff1}$) to the value from the spectroscopy\footnote{We find that if we choose different fixed parameters (e.g., fix $T_{\rm eff2}$ or $T_{\rm eff2}/T_{\rm eff1}$ instead, the derived $T_{\rm eff1}$ and $T_{\rm eff2}$ from the binary modelling are also similar to the spectral values, within 1.6 sigma ($\approx 100 K$). Thus, our results are not sensitive to the choices of fixed parameters. }. The fitting parameters include the temperature 
ratio $(T_{\rm eff2}/T_{\rm eff1}$), relative radii ($R_1/a, R_2/a$), eccentricity ($e$), argument of periastron ($\omega$), systematic velocity ($\gamma$), primary semi-velocity amplitude (pk), mass ratio ($q=M_2/M_1$), time of periastron passage($T_{peri}$), and orbital inclination $(i)$. We fixed the orbital period to the value obtained in Balona (2018): $P=13.9958015$d.
We first assume pseudo-synchronous rotation and find the resulting model $v \sin i$ are larger than the observed $v \sin i$ from 
spectroscopy. We then change the stellar rotation period ($P_{rot}$) to values that are consistent with spectroscopic $v\sin i$ ($P_{rot}=16.7$d, about $1/5$ of the pseudo-synchronous rate\footnote{In eccentric binaries, Hut (1981) showed that, when averaged over a long timescale, the stellar spins reach to an equilibrium state with the pseudo-synchronous rotation so that no average tidal torque is exerted on either star.}) and redo the fit. The phase-folded Kepler light curves and the radial velocities are shown in Figure 2, with the best-fitting LC and RV models in solid lines.
Table 2 contains the binary model parameters.
Note that the $\log g$ values from the binary model are somewhat different from the spectroscopic values. It is well known that $\log g$ cannot be determined to high precision, and also the spectral lines used in this Echelle order are not very sensitive of pressure broadening.

\section{Evolutionary Stage}

Our best binary model suggests the two stars have almost the same mass, $M_1=M_2=2.05M_{\odot}$ with a $1 \sigma$ error of 0.03, but very different radii ($R_1=2.96, R_2=2.51R_{\odot}$). This may seem to be surprising as the two stars must be coeval. It turns out the two stars have evolved to late stages of the main sequence when the stellar radius changes rapidly. A slight mass difference can result in a large radius difference. This can be seen in the isochrone plot of Figure 3. It shows isochrones from the Yonsei-Yale model (Yi et al.\ 2001) with stellar ages of 0.9, 1.0, and 1.1 Gyr. All these isochrones have solar-metallicity with chemical mixtures of Grevesse \& Noels (1993). On the 1.0 Gyr isochrone, we marked four points with their corresponding mass and radius labeled in the legend.  A mass difference of $\Delta M\approx 0.07M_\odot$ can result in a radius difference of $\Delta R \approx 0.7 R_{\odot}$. Thus the observed $\log g$ and $T_{\rm eff}$ of the two stars are, within one sigma, in agreement with predictions of two coeval stellar structure models. In the next section, we solidify our conclusion on the evolution stage by using asteroseismology of both p and g modes.

\section{Interpretation of Pulsations}

After subtracting the binary light curve, we performed a Fourier analysis of the residuals with the Period04 (Lenz \& Berger 2005). The Fourier amplitude spectrum is shown in Figure 4.  We extract significant frequencies (listed in Table 4 and 5, see below) by using a standard pre-whitening procedure.
The low-frequency regime ($f < 5$ d$^{-1}$) has very dense pulsation modes which are mostly self-excited $\gamma$ Dor type g modes, typical for an evolved A-star. We also identified oscillations that are likely tidally excited (see below).
In the high-frequency regime, most of the oscillations are located in the range from 15 to 18 day$^{-1}$. These are typical pressure modes of $\delta$ Scuti type.

\subsection{$\delta$ Scuti type p modes}

$\delta$ Sct stars are fast-rotating main sequence (MS) and post-MS stars with masses from $1.5$ to $2.5M_{\odot}$ and effective temperatures ($T_{\rm eff}$) from about $6500K$ to $9000K$. The majority of {\it Kepler} $\delta$ Sct stars pulsate with frequencies in the range of $10-30$ day$^{-1}$ (Bowman \& Kurtz 2018). Only very young $\delta$ Scts can pulsate at frequencies $>40$ day$^{-1}$ and can be as high as 70 day$^{-1}$.
The unstable modes shift to lower frequencies as $T_{\rm eff}$ decreases, and this can be understood as the partial ionization zone moving to the inner region where the local thermal timescale is longer and thus the comparable pulsation frequencies are lower (Pamyatnykh 1999). Observationally, Barcel{\'o} Forteza et al.\ (2018) established this $T_{\rm eff}-\nu_{max}$ relation for {\it CoRoT} and {\it Kepler} $\delta$ Scts. The stellar density also decreases, so do the radial mode frequencies and the frequency spacing between the adjacent radial modes (referred to as large frequency separation).

In KIC 4142768, the observed p-mode frequencies are located at about $15-18$ day$^{-1}$ (Figure 4). The left panel of Figure 5 is the HR diagram for a stellar model with $M=2.05M_{\odot}$ calculated with the MESA evolution code (Paxton et al.\ 2011, 2013). We adopt the solar metallicity with the {\it gs98}  chemical mixtures (Grevesse \& Sauval 1998) and a fixed initial helium abundance of $Y=0.28$. The stellar models have a fixed mixing-length parameter of $\alpha=1.8$ with the convective treatment of Henyey et al.\ (1965). Four evolutionary stages from the Zero Age Main Sequence (ZAMS) to Terminal Age Main Sequence (TAMS) are labeled (A, B, C, D). The $\pm 2\sigma$ credible region of the observed radius of the primary (secondary) star is represented by the green (blue) line. In the right panel, we show the mode stability parameter $\eta$ as a function of p-mode frequencies. The calculation of p-mode stabilities was performed by using Dziembowski's non-adiabatic code NADROT (Dziembowski 1971, 1977). From stage A to stage D, the excited p-modes frequencies (the top of the hill) shift from 50 to 15 d$^{-1}$. Compared to the observed p-mode frequencies, we find that model D matches KIC 4142768, and model C, B, A cannot excite p modes in the observed frequency range.

Regular frequency spacings have been observed in $\delta$ Sct stars. This is especially obvious if using a large sample of stars (Michel et al.\ 2017; Paparo et al.\ 2016).
This empirical frequency separation is similar to large separation and is found to be proportional to the square root of mean stellar density when calibrating with well-measured mass and radius in eclipsing binaries and interferometry (Garcia Hernandez et al.\ 2015).
The observed p-mode frequencies in KIC 4142768, as shown in detail in the lower panel of Figure 4, seem to form two clusters, with a spacing of about $2.5-3$ d$^{-1}$.
This kind of clustering and regular frequency spacing patterns have been found in $\delta$ Sct stars (e. g., Maceroni et al.\ 2014). And it can be explained by trapped non-radial modes clustering around the closest radial modes (Breger et al.\ 2009; Dziembowski \& Krolikowska 1990). Thus the spacing between the clusters is usually interpreted as the large frequency separation. Comparing to the theoretical large frequency separation in the right panel of Figure 5 (the short solid-line ended with two arrows), which can be regarded as the spacing between adjacent radial modes (black circles), we can see this value decreases from stage A to D. Model D has a spacing of about 3 d$^{-1}$, which best matches the observed spacing of KIC 4142768. The frequency spacings of Model A, B, and C ($\approx 5$, 4.5, and 4.0 d$^{-1}$, respectively) are too large.

$\delta$ Sct pulsators are excellent clocks to perform the time delay analysis (Hulse \& Taylor 1975; Shibahashi \& Kurtz 2012; Telting et al.\ 2012). Orbital parameters and the origin of pulsations can be found in this way (Murphy et al.\ 2014, 2016, 2018; Schmid et al.\ 2015). However, the large uncertainties in the time delay measurements ($\sim 20s-80s$) prevent us from yielding any conclusive result for such a close orbit ($80$ light-seconds across) of KIC 4142768.

\subsection{Tidally excited g modes}

In Figure 4, a remarkable feature in the g-mode regime ($f \le 5 $ d$^{-1}$) is a series of peaks at orbital harmonics $N f_{orb}=N \times 0.07145$ d$^{-1}$. These integer multiples of the orbital frequency are labeled with the vertical dotted lines. We ascribe these peaks to two origins as described below.

The precise photometry of Kepler poses challenges in the modeling of eclipsing binary light curves. Even state-of-the-art modeling tools still fail to model the observations of heartbeat stars perfectly. For example, as shown in Hambleton et al.\ (2016) and Welsh et al.\ (2011), the light curve residuals are not gaussian-like and still show systematic variations.
The imperfect removal of the eclipsing binary light curve thus can generate alias peaks of the form $Nf_{orb}$, and these peaks should have low amplitudes. We ascribe the consecutive low-amplitude $Nf_{orb}$ peaks in the range of 1.5 to 3 d$^{-1}$ to this imperfect removal. 

Some $Nf_{orb}$ peaks, especially in the range of 0.5 to 1.5 d$^{-1}$ (N $\approx 8$ to $20$), have very high amplitudes, and cannot be explained by the imperfect removal of the binary light curve. These peaks are most likely tidally excited oscillations (TEOs). 
The removal-generated $Nf_{orb}$ peaks should also be present in this frequency range, but their amplitudes are much lower than the observed amplitudes here in this star. In the TEO scenario, the tidal potential from the companion star can be decomposed spatially into spherical harmonics, and couples with another set of  spherical harmonics describing the star's gravitational potential perturbation of intrinsic eigenmodes, and the most effective coupling happens to the $l=2$ g modes. Temporally, the tidal potential can be decomposed into Fourier series and each component has a driving frequency $N f_{orb}$ (with $N$ from zero to infinity). When a driving frequency is close to an eigen-mode frequency of the star, the mode can be excited to a large amplitude and thus produce a large temperature perturbation. Finally, the tidal response reveals itself in the light curve as luminosity perturbations (Burkart et al.\ 2012; Fuller 2017). 

To safely identify TEOs, we are being conservative and adopt a more strict criterion for significant frequencies ($S/N \ge 10$) as opposed to the traditional $S/N \ge 4$. The red solid line in Figure 4 indicates our noise model. With this criterion, all the $Nf_{orb}$ peaks from the imperfect binary removal are discarded, although we cannot rule out the possibility that certain orbital harmonics with $10 \ge S/N \ge 4$ are actually real TEOs.
The adopted significant TEOs are labeled with grey squares and red vertical lines in Figure 4.

To model these TEOs, we first evolve a star with the observed parameters of the primary ($M=2.05$, $R=2.96R_{\odot}, Z=0.02$) with the MESA evolution code. We adopt the OPAL opacity table and {\it gs98} composition (Grevesse \& Sauval 1998). Then we calculate the non-adiabatic eigen-frequencies and eigen-functions of the star for $l=2, m=(2,0,-2)$ modes with the GYRE oscillation code (Townsend \& Teitler 2013). The rotation is included in the traditional approximation. We then use these free-oscillation eigenfunctions as basis vectors to construct the tidal response of the star following the treatment in Fuller (2017).
The resulting flux variation ($\Delta L/L$) from the dynamical tide (after subtracting the equilibrium tide component from the full stellar response) is shown in Figure 6 (Diamonds). The $\Delta L/L$ sensitively depends on the resonance detuning parameter\footnote{The difference between a driving frequency ($Nf_{orb}$) and the closest intrinsic eigenmode frequency.}, for which we cannot determine accurately due to observational and modeling uncertainties. We can instead assume the detuning parameter is a random variable, uniformly distributed around its minimum value ($=0$) and maximum value ($=$ half of the g mode frequency spacing at a forcing frequency $Nf_{orb}$). Thus we can calculate the corresponding $\Delta L/L$ statistically\footnote{Section 4 in Fuller (2017)}. The blue shaded region indicates the $\pm 2 \sigma$ credible region of $\Delta L/L$.  The observed TEOs are shown as grey squares. The theory indeed predicts expected large TEOs at $N$ about $10-20$, in agreement with the observations. There is an oscillation observed at the $N=3$ orbital harmonic that is not predicted by our TEO modeling. Rossby modes have been observed in A-, B-stars as well as heartbeat binaries (Saio et al.\ 2018; Li et al. 2019b), although the mechanism of excitation is not investigated. Tides could excite rossby modes at low frequencies. However, the $N=3$ pulsation in this system has a frequency higher than twice the inferred rotation frequency, so it cannot be an $m=2$ Rossby mode. It is thus more likely to be an artifact from the data reduction. Similar artificial low-frequency harmonics have been reported in other tidally oscillating heartbeat stars as well (Pablo et al.\ 2017).

Non-linear effects can generate tidal oscillations that are not orbital harmonics. They are usually in the form of daughter modes, satisfying mode resonance conditions: $f_b+f_c \approx f_a=Nf_{orb}$ (Weinberg et al.\ 2012; O'Leary \& Burkart 2014). We did not consider this effect here since such combinations are not found in this binary.

We also model the phases of the flux variations as they convey important information on the mode identification (Burkart et al.\ 2012; O'Leary \& Burkart 2014; Guo et al.\ 2017b). Simply speaking, the phases of TEOs only depend on the geometric orientation of the star and the coordinate of the observer in the binary system. With reasonable assumptions such as mode adiabaticity and spin-orbit alignment, the TEO phases can be expressed as a function of the argument of periastron of the binary orbit ($\omega_p$). For $m=0$ modes, the phases are at 0.25 or 0.75; and for $m=2$ modes, the phases are related to $\omega_p$ by $\delta_N=0.25+m[0.25-\omega_p/(2\pi)]$. 
Given $w_p=328^{\circ}$ for KIC 4142768, we derive the adiabatic phases $\delta_N=$ 0.07 or 0.57 for ($l=2,m=2$) modes.
In Figure 7, we show these simple `theoretical adiabatic phases' as red dashed lines. The observed TEO phases are indeed distributed around these two lines (with some scatter), so they are consistent with our expectation that they are tidally excited $m=2$ modes.. 

Detailed modeling of TEO phases requires the inclusion of mode non-adiabaticity. The non-adiabatic effect will add a phase shift to the flux variation of TEOs. In Figure 7, blue circles indicate the theoretical phases of $\Delta L/L$ from the detailed modeling with non-adiabatic calculations following Fuller (2017). Note that the observed scattering of TEO phases around the adiabatic values is about $0.05-0.1$, our calculation indeed can reproduce a scattering at this level. At low frequencies ($N \lesssim 8$), the tidally excited g modes have high radial orders. These modes couple weakly with the tidal potential and suffer much larger non-adiabatic effect. For these reasons, their amplitudes are smaller and their phases deviate strongly from the adiabatic prediction.

\subsection{$\gamma$ Doradus type g modes}

$\gamma$ Dor stars are F- or A-type dwarfs with masses from 1.3 to 2.0 $M_{\odot}$. They are characterized by low-frequency g-mode pulsations with periods ranging from about 0.3 to 3 days. Ouazzani et al.\ (2018) and Mombarg et al.\ (2019) found that fast-rotating $\gamma$ Dor stars are younger and less massive than the slow rotators. It is expected that high-mass $\gamma$ Dor stars are likely to be slow rotators since they are closer to the TAMS than low-mass stars. KIC 4142768 is indeed a high-mass ($M=2.05M_{\odot}$) and evolved $\gamma$ Dor star with a slow rotation rate ($v \sin i \approx 8$ km s$^{-1}$).

After masking all the orbital harmonic frequencies, we show the Fourier spectrum in the g-mode regime in the upper panel of Figure 8, with the horizontal axis in period ($P$). A notable feature is that within 1.4 and 2 day$^{-1}$, the peaks are near-equally spaced with a typical period spacing ($\Delta P$) of about $3000$ seconds. High-order g modes are expected to be equidistant in period ($P$), with a typical period spacing $\Delta\Pi_{l}$. 
The $P$ vs $\Delta P$ diagram of KIC 4142768 is shown in the lower panel of Figure 8, and the black rectangle highlights the region where the regular period spacings are most remarkable with $\Delta P$ from about 2500s to 3500s. These are most likely $l=1$ modes since $l=2$ modes should have lower spacings.  Only peaks with small period spacings ($\approx 1800 $ seconds) in the short period region of the Fourier spectrum ($P<1.2$ d) are likely to be $l=2$ modes (or trapped $l=1$ modes, see below). We labeled all the peaks we used to produce the $P-\Delta P$ diagram with vertical dotted lines and they are listed in Table 5. 

At an orbital inclination of $76$ degrees, we expect that the axisymmetric modes ($m=0$) should have lower amplitudes than sectoral modes ($m=\pm 1$), assuming the spin and orbital axes are aligned. Previous studies also found that prograde dipole modes are more dominant in {\it Kepler} $\gamma$ Dor stars (Van Reeth et al.\ 2016, Li et al.\ 2019a,b).
Assuming the near equally-spaced g modes in the rectangle are $l=1,m=1$ modes, we can fit the $P$ vs $\Delta P$ with the asymptotic relations for high-order g modes: $P_{nl, co}\approx \Delta \Pi_l(n+0.5)$ (after transforming the frequencies/periods to the inertial frame: $f_{in} = f_{co} + mf_{rot}$), where $\Delta \Pi_l=\Delta \Pi_0/\sqrt{l(l+1)}$. This relation can be extended when rotation is included following the traditional approximation (Unno et al.\ 1989; Bildsten et al.\ 1996), with $\sqrt{l(l+1)}$ replaced by the eigenvalue of the Laplace tidal equations $\lambda$. The slope and vertical displacement of the $P$ vs. $\Delta P$ diagram can provide us information on the mode identification and the near-core rotation rate ($\Omega_{\rm core}$). Internal rotation rates of many $\gamma$ Dor and Slowly Pulsating B-stars (SPB) have been measured (Van Reeth et al.\ 2016; Li et al.\ 2019a,b). The flat $P-\Delta P$ of KIC 4142768 suggests that the near-core region of the primary star is rotating slowly.   The final fit is shown as the red solid and dotted lines in the rectangle as we only choose the most well-behaved region in the $P$ vs $\Delta P$ diagram. Our exercise here for KIC 4142768 yields a near core rotation rate\footnote{The method of deriving core-rotation from the $P-\Delta P$ diagram is not very sensitive when the rotation rate is very low. The error bars here are underestimated and we thus only take this value as evidence for slow rotation.} $\Omega_{\rm core} = 0.006\pm 0.003$ day$^{-1}$ and a dipole mode asymptotic period spacing of $\Delta \Pi_{l=1}=3040 \pm 18$s (or $\Delta \Pi_0 = 4300 \pm 25$s). The value of $\Delta \Pi_{l=1}$ is roughly in agreement with the representative models B and C, which have $\Delta \Pi_{l=1}$ of  $3099$s and $2912$s, respectively (Figure 9). Note that this combination of $(\Omega_{\rm core}$ and $\Delta \Pi_{0}$) can also fit the observed $P$ vs $\Delta P$ pattern in the low-period region ($P<1.2$ d), assuming that they are $l=2, m=2$ modes (e.g, green circles in Figure 8). We emphasize here the asymptotic period spacing $\Delta \Pi$ derived from that $P-\Delta P$ diagram suffers from systematic uncertainties if the period spacing pattern is not well observed. This is especially true for evolved stars, where trapped modes form many dips in the $P$ vs $\Delta P$ diagram. If we choose to use more red data points in Figure 8, instead of just those inside the rectangle, we can derive a smaller $\Delta \Pi_{l=1} \approx 2500s$ (and a similar slow near-core rotation rate $\approx 0.01$ day$^{-1}$), in better agreement with model C and D ($\Delta \Pi_{l=1}=2912, 2530s$, respectively).

To better compare the observed g-modes periods with the theory, we calculate the individual g-mode frequencies for the four models with different evolutionary stages (A, B, C, D in Figure 5). We assume a solid-body rotation and adopt a rotation rate of $V_{eq}$ based on spectroscopic $v\sin i$. The periods and period spacings of the prograde sectoral $l=1$ and $l=2$ modes are shown in Figure 8. The asymptotic period spacings are indicated by the red dotted line. It can be seen that prograde dipole modes of model C and D best match the observed g-mode period spacings. 

Our calculations do not include near-core mixing, e.g., convective-core overshooting parameterized by $f_{ov}$ or the diffusion in the envelope ($D_{\rm diff}$). This mixing will smooth the $P$ vs $\Delta P$ pattern (Bouabid et al.\ 2013) and observations of $\gamma$ Dor seem to favor a moderate level of overshooting with $f_{ov}\approx 0.015$. It is possible that a certain level of mixing could improve the modeling of the observed g modes.

\section{Comments on Balona (2018)}
Our interpretation of the tidal effect on pulsations differs from those in Balona (2018), hereafter B18. We did not find combination frequencies of the form $f \pm Nf_{orb}$. B18 subtracted a `heartbeat light curve' in the phase-folded space (his Figure 5) before performing the frequency analysis. His `heartbeat light curve' contains many orbital harmonics pulsations. We suspect his treatment of removal induces some modulations of the p modes which can explain why B18 found many peaks in the form of $f \pm Nf_{orb}$. B18 interpreted these peaks as tidally excited splittings based on the theory of Reyniers \& Smeyers (2003), which is based on the assumption of circularized and synchronized binary. B18 demonstrated a RV orbit based on the preliminary result of Guo (2016). We have updated the RV measurements in this paper, and the resulting orbit is more eccentric, with a much higher eccentricity $e \approx 0.6$. Thus the theory of Reyniers \& Smeyers is no longer applicable. Note that there are indeed observational evidence of tidal splittings in binaries with circular orbits and synchronized components (e.g., Guo et al.\ 2016; Handler et al.\ 2019), though we see no evidence for tidal splitting here.

\section{Discussion}
As shown in Figure 9, less evolved models tend to show more obvious period spacing patterns. For evolved models, the $P-\Delta P$ diagram shows numerous dips due to mode trapping in the chemical gradient region near the convective-core boundary. It is thus more challenging to discern a reliable pattern from the observed g modes (see Figure 9). Although we find reasonable agreement between theory and observations in terms of g-mode asymptotic period spacings for this evolved system, it is more desirable to perform a multi-dimensional search (e.g., mass, age, metallicity, overshooting, etc.) for the best stellar model to match the observed traditional observables (M, R, $T_{\rm eff}$, [Fe/H]) and seismic observables (Moravveji et al.\ 2015; Aerts et al.\ 2018, Schmid \& Aerts 2016; Mombarg et al.\ 2019). Being a very slow rotator, the effect of rotation can be satisfactorily accounted for by the perturbative approach.

In addition, we can extend our spectroscopic analysis to KIC 4142768 and measure the abundances of individual elements. Combined with kinematic information from GAIA, it is possible to characterize the formation history of this binary and its relation to nearby stars.

It is likely that the primary star is a hybrid pulsator showing both the $\delta$ Scuti-type p modes and the tidally excited and self-driven $\gamma$ Dor type g modes. However, we cannot rule out the possibility that some of the $\gamma$ Dor type g modes are from the secondary. 
Our derived fundamental stellar properties of KIC 4142768 consistently explain all the observations:  the unstable $\delta$ Sct type p modes and their regularities, the $\gamma$ Dor type g modes and their period spacings, and the amplitudes and phases of tidally excited g modes. This demonstrates the advantage of studying hybrid pulsating stars in eclipsing binaries systems. Gaulme \& Guzik (2019) identified $303$ pulsating EBs in about 3000 {\it Kepler} EBs. TESS 2-min cadence data can yield about $300$ eclipsing binaries per sector, and more than $1/10$ (conservative estimation) are expected to contain pulsating stars. We are just beginning to scratch the surface of the observed pulsating EBs, and the hybrid pulsating nature is still rarely exploited. We also expect to have many B-type $\beta$ Cephei/SPB hybrid pulsators (Handler et al.\ 2009; Pedersen et al.\ 2019).  Hybrid p- and g-mode pulsations have also been found in sub-dwarf B-stars (sdB) (Baran et al.\ 2011, 2017; Reed et al.\ 2010, 2019) and proto-Helium Extremely Low Mass White Dwarfs (Maxted et al. 2013).

Thanks to the probing capability of g modes and mixed modes, an increasing number of stars have both the surface and near-core rotation rates measured (Salmon et al.\ 2017; Aerts et al.\ 2017, 2019). It reveals the angular momentum transfer history through the life of stars from the main-sequence to the giant branch. However, among the aforementioned stars, very few are binaries. Previously (Guo et al.\ 2017a; Guo \& Li 2019), we find the short-period eclipsing binary KIC 9592855 and KIC 7385478 both contain a $\gamma$ Dor pulsator that is synchronized at the surface and the near-core region. The eccentric binary KIC 4142768 has a slow-rotating core and a sub-pseudo-synchronous slow-rotating surface. More studies like these two will help us to calibrate the timescale of tidal circularization/synchronization and angular momentum transfer inside stars.

 
\acknowledgments 
We are in debt to the referee whose suggestions improve the quality of this paper.
We thank the {\it Kepler} team for making such exquisite data publicly available, and the MESA team for developing excellent modeling tools. ZG thanks Jerry Orosz for making the ELC code available. The usage of programs developed by the team of Douglas Gies in Georgia State University is acknowledged.
This work is partially supported by the Polish NCN grants grant 2015/18/A/ST9/00578.
JF acknowledges support by an
Innovator Grant from The Rose Hills Foundation and
the Sloan Foundation through grant FG-2018-10515. KH, JF and AS acknowledge support through NASA ADAP grant (16-ADAP16-0201).
The authors wish to recognize and acknowledge the
very significant cultural role and reverence that the summit of Mauna Kea has always had within the indigenous Hawaiian
community. We are most fortunate to have the opportunity to
conduct observations from this mountain.



\clearpage



\begin{figure} 
\begin{center} 
{\includegraphics[angle=90,height=12cm]{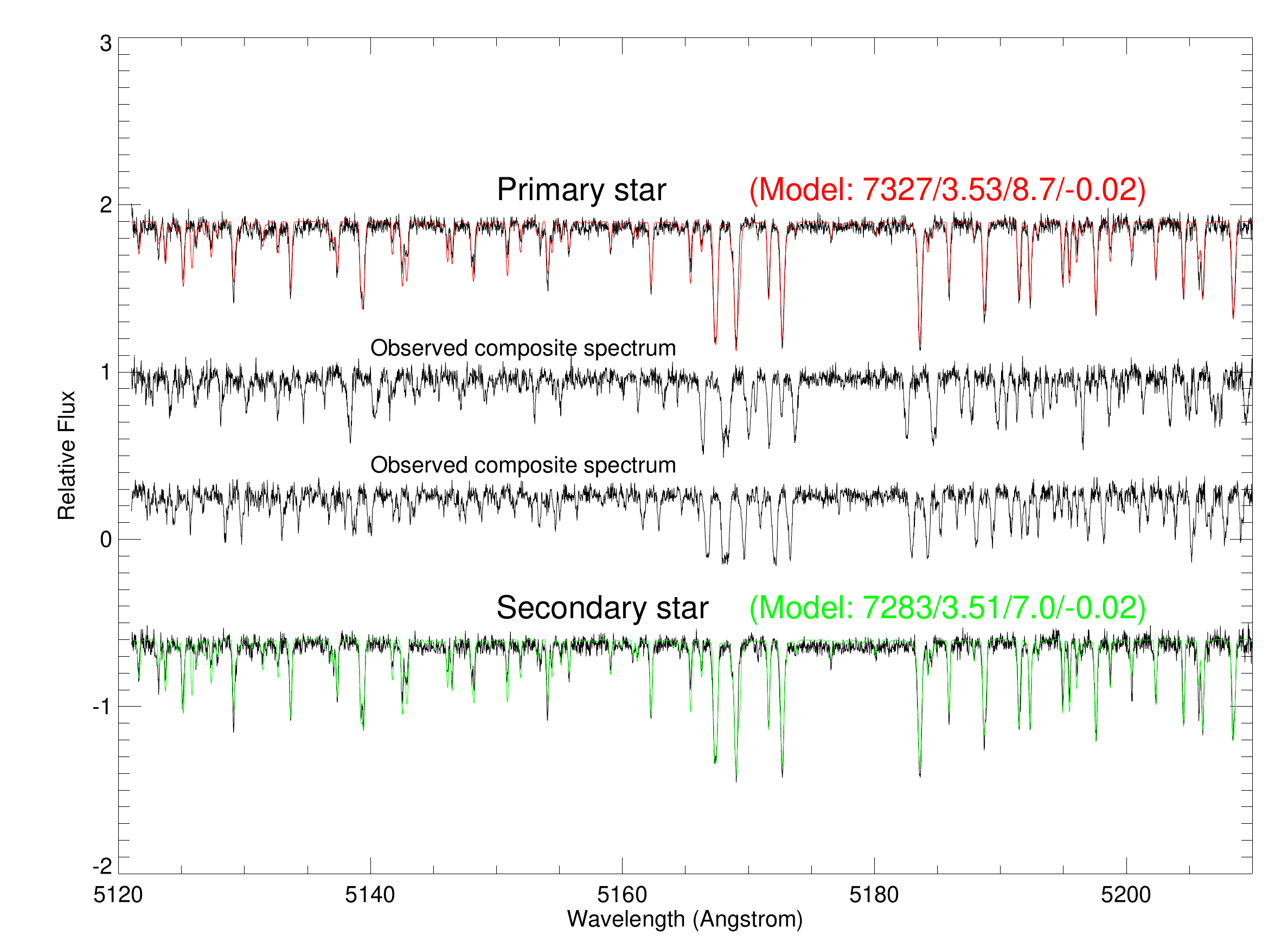}} 
\end{center} 
\caption{Observed composite spectra (middle two) and the disentangled indiviadual spectra of the primary and secondary (upper and lower). The red/green solid line reprsents the best-fitting BLUERED model for the primary/secondary star. The atmospheric parmeters ($T_{\rm eff}/\log g/v \sin i/[Fe/H]$) are labeled (Table 1).}
\end{figure} 

\begin{figure} 
\begin{center} 
{\includegraphics[angle=0,height=12cm]{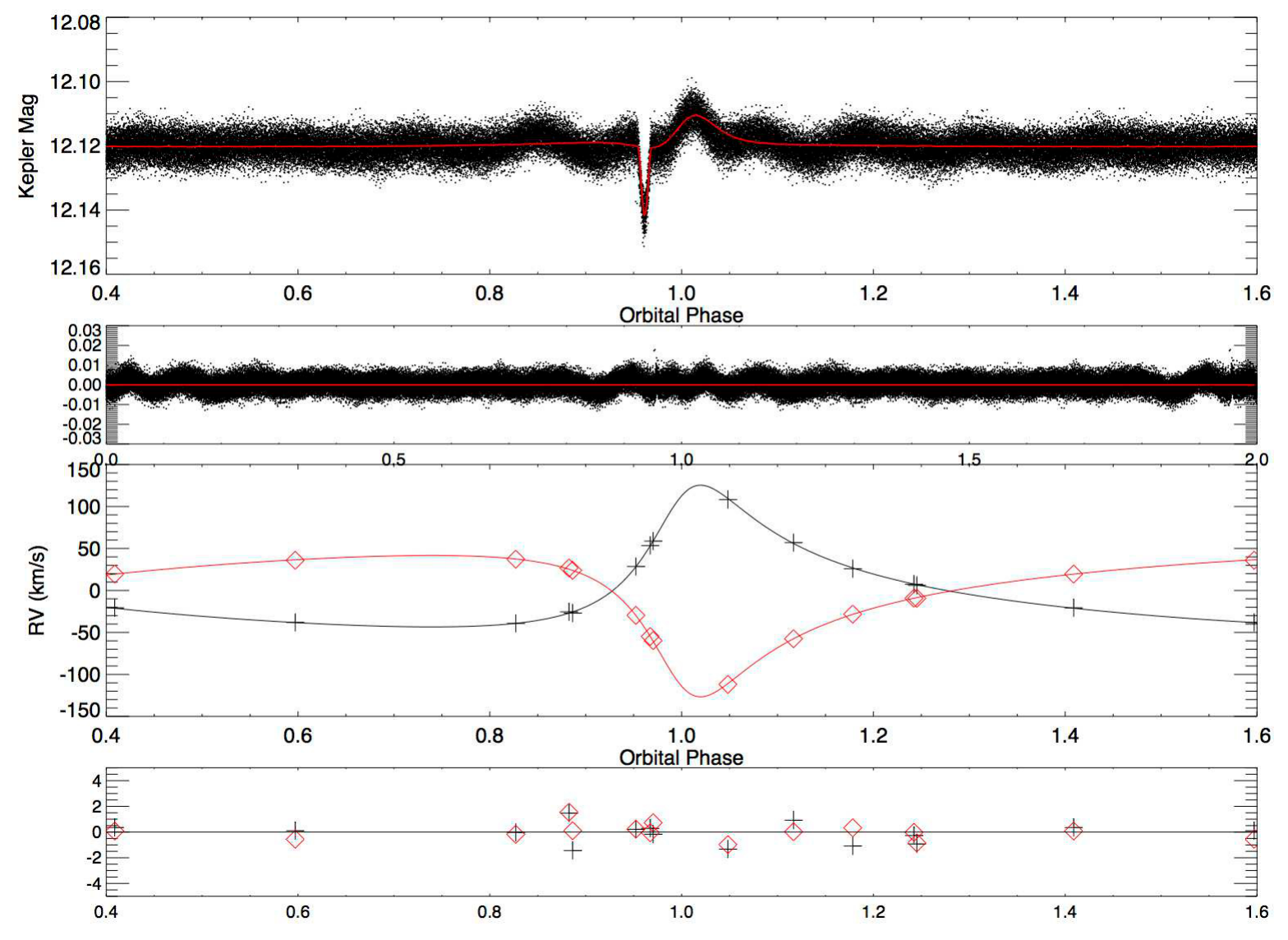}} 
\end{center} 
\caption{The phase-folded {\it Kepler} light curve (upper) and radial velocity curve (lower) with the best-fitting models from the ELC (solid lines) overploted. The two narrow panels show the corresponding residuals. Note that the light curve residuals clearly show signatures of orbital harmonic pulsations. }
\end{figure}

\begin{figure} 
\begin{center} 
{\includegraphics[angle=0,height=12cm]{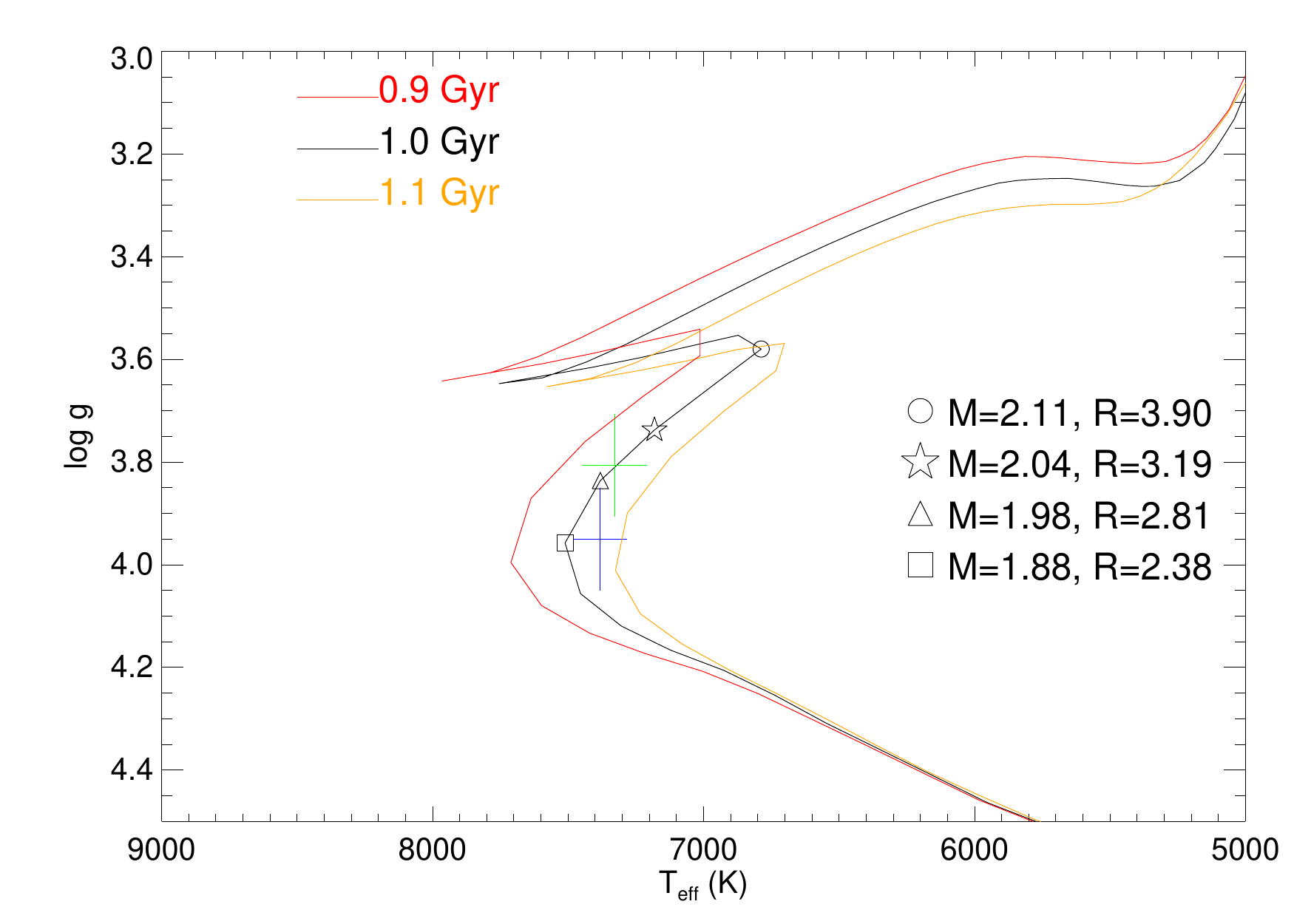}} 
\end{center} 
\caption{The derived $\log g$ and $T_{\rm eff}$ of KIC 4142768 and the isochrones of $1.1, 1.0, 0.9$ Gyr with solar metallicity from the Yonsei-Yale evolution models (red, black, orange, respectively). Four locations on the 1.0 Gyr isochrone are marked and the corresponding masses and radii are labeled. A small mass difference of $\Delta M \approx 0.07M_{\odot}$ can result in a large radius difference of $\Delta R \approx 0.7R_{\odot}$. }
\end{figure}

\begin{figure} 
\begin{center} 
{\includegraphics[angle=0,height=14cm]{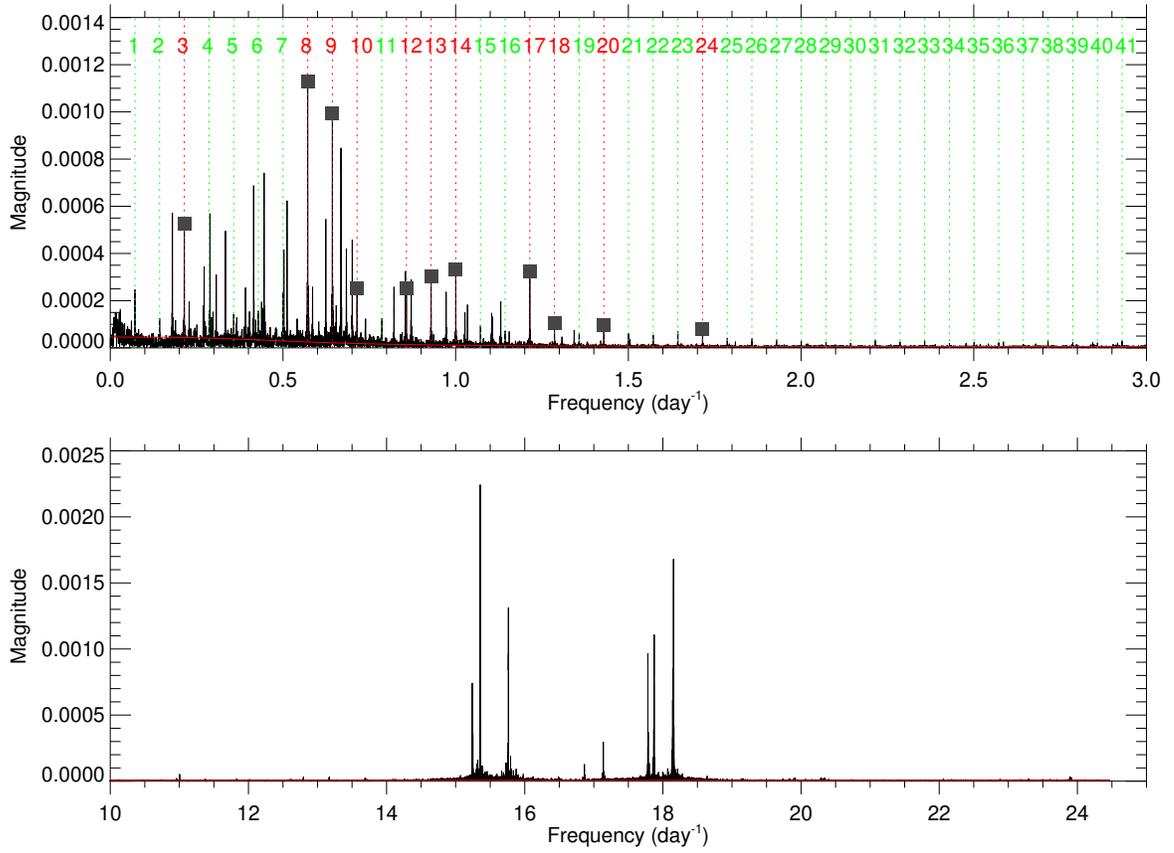}} 
\end{center} 
\caption{The Fourier amplitude spectrum of light curve residuals after subtracting the binary light curve. The upper and lower panels show the g- and p-mode regions, respectively. Integer multiples of orbital frequency are labeled by the dotted vertical lines, with those identified as TEOs in red (with $S/N > 10$) and those likely arising from the imperfect binary light curve removal in green. }
\end{figure}

\begin{figure} 
\begin{center} 
{\includegraphics[angle=0,height=10cm]{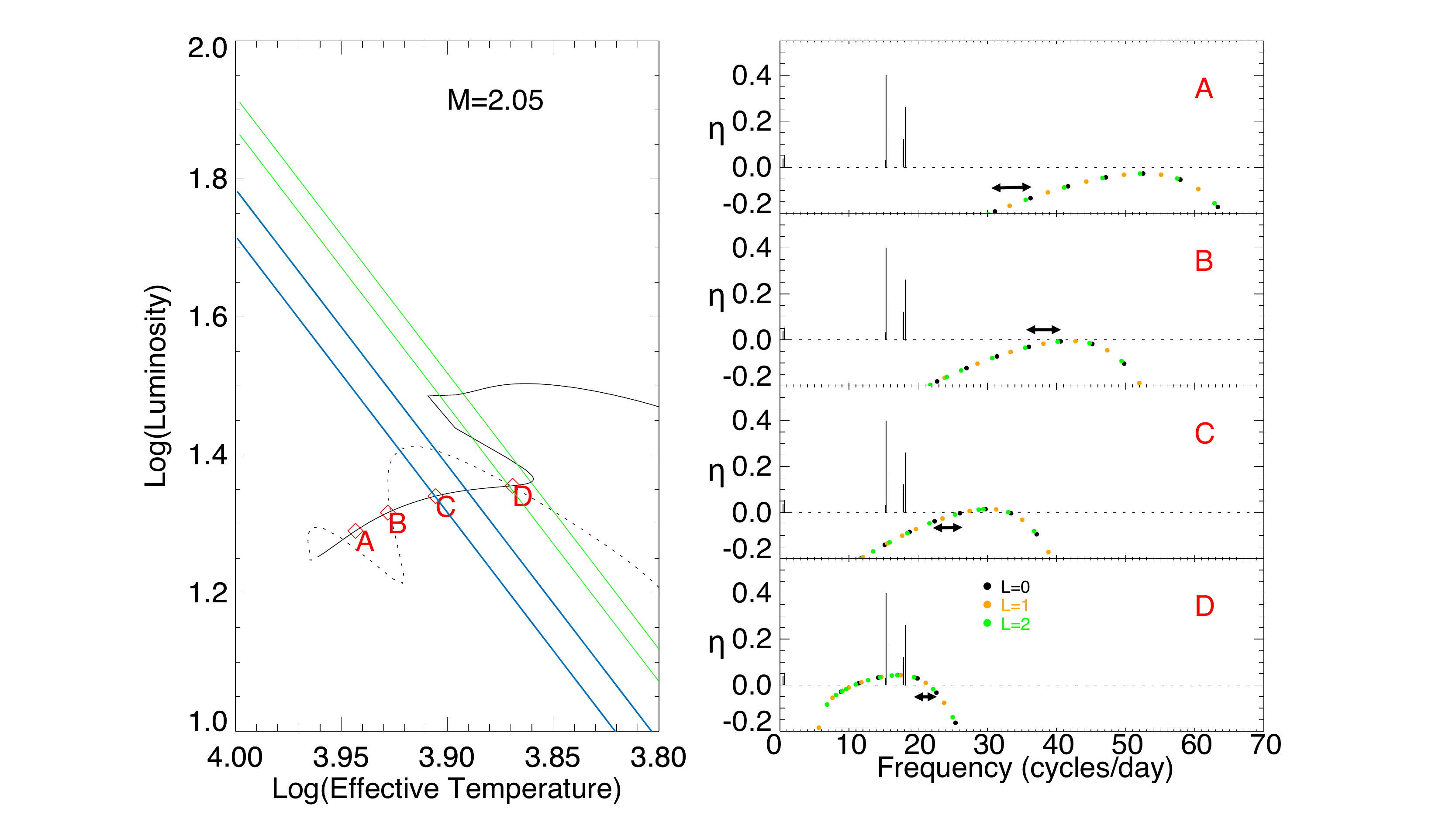}} 
\end{center} 
\caption{Left panel shows the evolutionary track of a MESA stellar model with $M=2.05M_{\odot}, Z=0.02, f_{ov}=0.0$. The dotted and solid lines represent the pre-MS and MS phases, respectively. Four representative evolution stages are labeled as A, B, C, and D.  The two strips represent the iso-radius lines within two sigma of the observed $R_1=2.96R_{\odot}$ (green) and $R_2=2.51R_{\odot}$ (blue). The theoretical pulsation frequencies and instability parameters of these four stages are plotted in the right panel. The observed p-mode frequencies are scaled and overploted. }
\end{figure}

\begin{figure} 
\begin{center} 
{\includegraphics[angle=0,height=12cm,width=16cm]{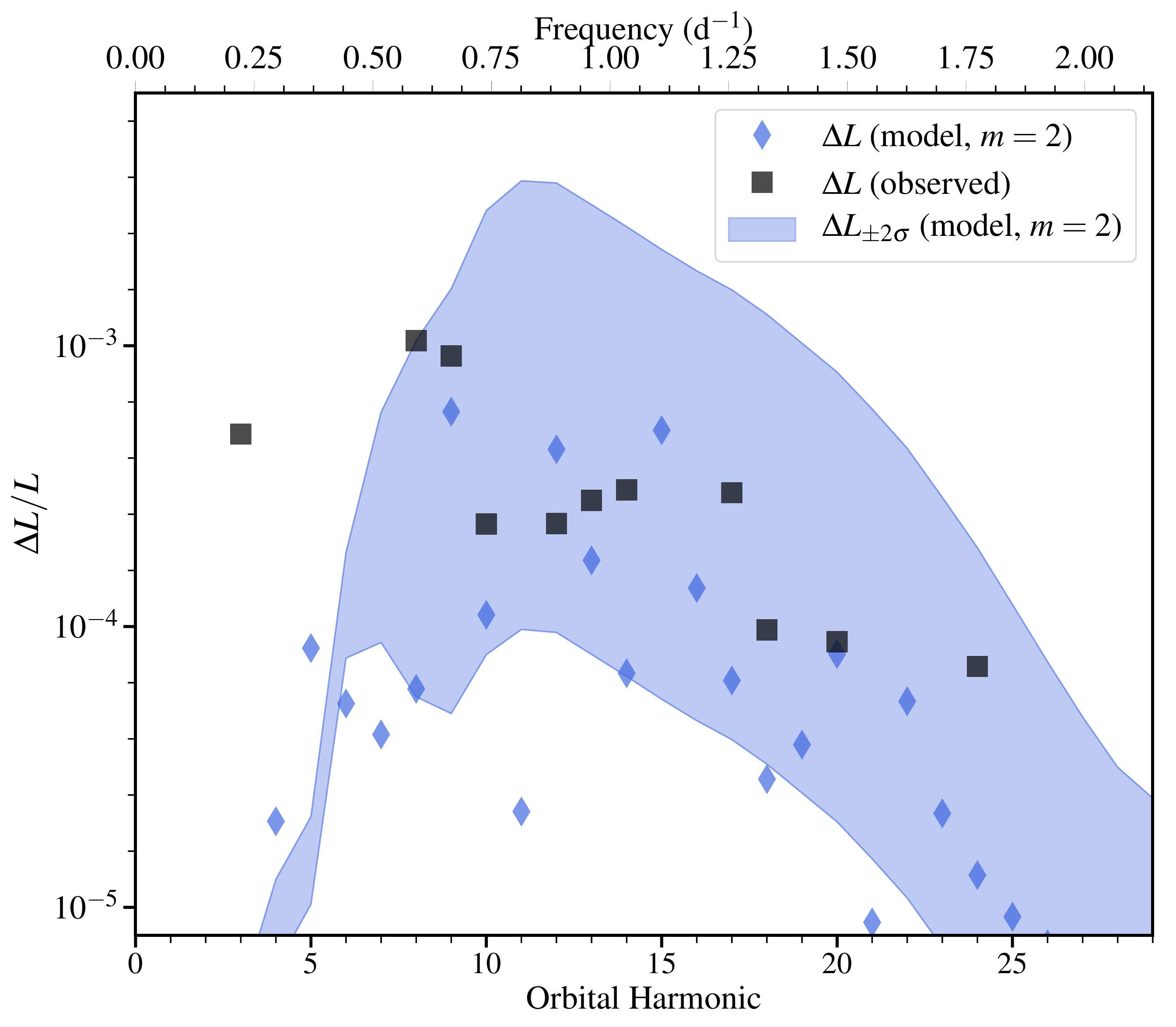}} 
\end{center} 
\caption{The modeling of pulsation amplitudes of TEOs. \textbf{The calculated TEO amplitudes from the model representing the primary star for $l=2,m=2$ modes are shown as blue diamonds. By treating the detuning parameter as a uniformly distributed random variable, the corresponding $\pm 2 \sigma$ credible region of TEO amplitudes is indicated by the shaded region.} The observed amplitudes are represented by the dark squares. }
\end{figure} 

%
\begin{figure} 
\begin{center} 
{\includegraphics[angle=0,height=12cm,width=20cm]{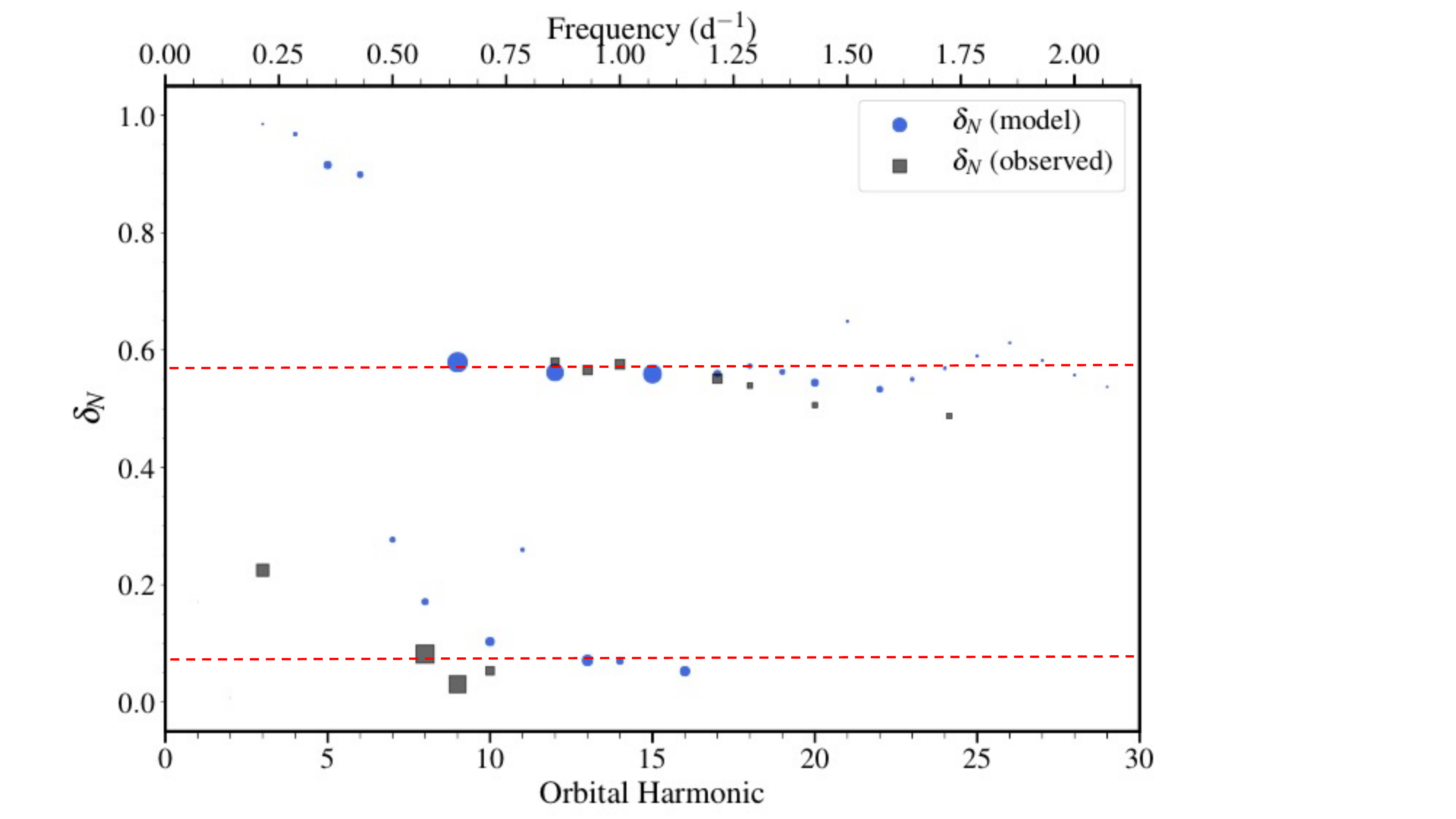}} 
\end{center} 
\caption{The TEO phases ($\delta_N$) of $l=2,m=2$ modes for each orbital harmonic. The observed and theoretical phases are represented by grey squares and blue circles, respectively, with the size of the symbols indicating the pulsation amplitudes. The two horizontal red lines ($\delta_N=0.07$ and 0.57) show the theoretical TEO phases from the adiabatic approximation.   }
\end{figure} 

\begin{figure} 
\begin{center} 
{\includegraphics[angle=0,height=14cm]{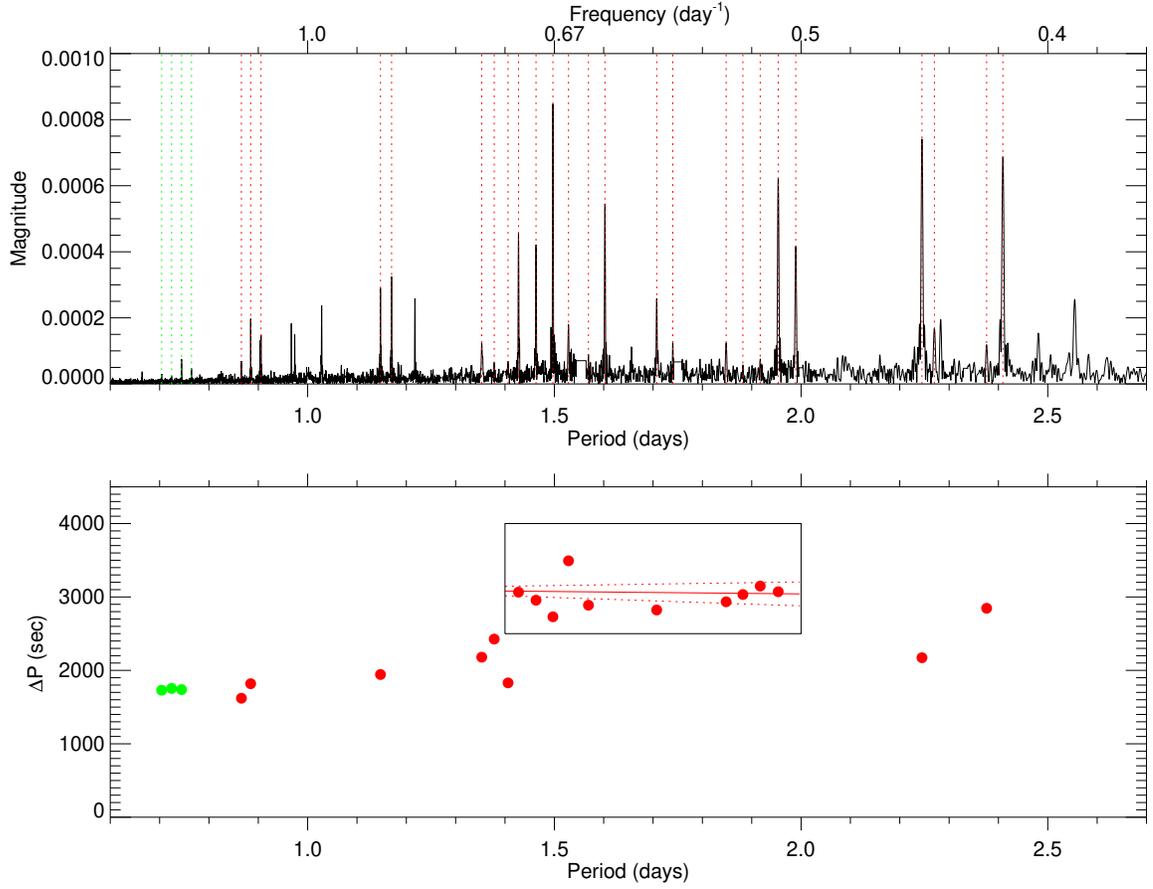}} 
\end{center} 
\caption{Identification of period spacing patterns in the g-mode regime. \textbf{Upper panel}: All orbital harmonics are masked and the \textbf{self-driven} g-mode oscillations are shown in period in the units of days. \textbf{Lower panel}: The period $(P)$ vs period spacing ($\Delta P$)  diagram. The rectangle highlights the g modes which show the most regular period spacing patterns. }
\end{figure}

\begin{figure} 
\begin{center} 
{\includegraphics[angle=0,height=14cm]{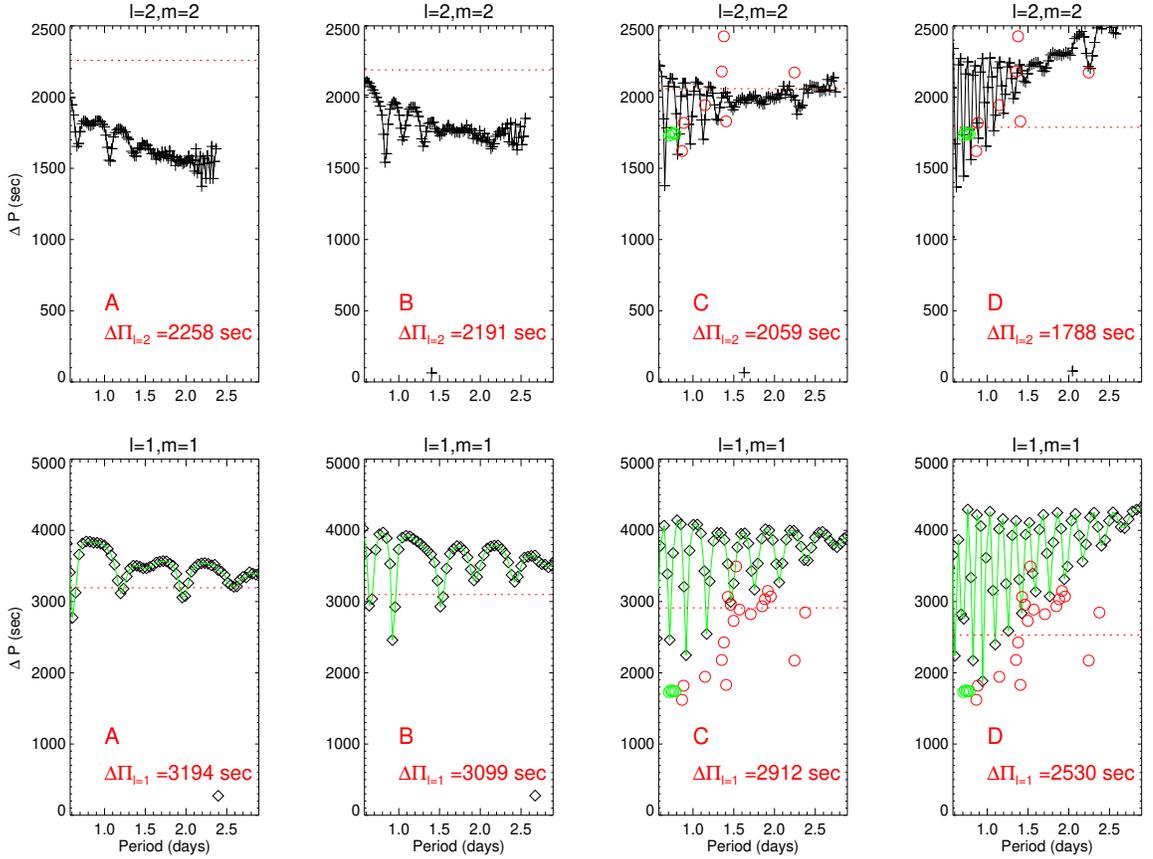}} 
\end{center} 
\caption{ \textbf{Upper panels}: Period vs Period Spacing ($\Delta P$) of ($l=2,m=2$) g modes calcualted with GYRE corresponding to the four evolutionary stages (A, B, C, D) in Figure 5; \textbf{Lower panels}: same plot but for ($l=1,m=1$) g modes. The observed period and period spacing of KIC 4142768 are overplotted as circles. The asymptotic period spacings for $\Delta\Pi_{l=1}$ and $\Delta\Pi_{l=2}$ are indicated by the red horizontal lines. These values decrease with stellar age. }
\end{figure}


\begin{deluxetable}{lccccc} 
\tabletypesize{\small} 
\tablewidth{0pc} 
\tablenum{1} 
\tablecaption{Atmospheric Parameters\label{tab1}} 
\tablehead{ 
\colhead{Parameter}   & 
\colhead{Primary star}     &
\colhead{Secondary star}     &
}
\startdata 
$T_{\rm eff}$ (K)               \dotfill & $7327 \pm 64$ & $7283 \pm 60$  \\ 
$\log g$ (cgs)  \dotfill & $3.53\pm 0.10$  & $3.51\pm 0.10$        \\ 
$v \sin i$ (km s$^{-1}$)     \dotfill & $8.7 \pm 0.2$        & $7.0 \pm 0.2$         \\ 
$\rm[Fe/H]$ \dotfill & $-0.02 \pm 0.05$        & $ -0.02 \pm 0.05$             \\   
\enddata 
\end{deluxetable} 

\clearpage

\begin{deluxetable}{lccc} 
\tabletypesize{\small} 
\tablewidth{0pc} 
\tablenum{2} 
\tablecaption{Binary Model Parameters\label{tab2}} 
\tablehead{ 
\colhead{Parameter}   & 
\colhead{Primary}      &
 \colhead{Secondary}  &
\colhead{System}       
}
\startdata 
Period (days) & & &$13.9958015 \tablenotemark{a}\pm 0.0000629$\\
Time of periastron passage, $T_{peri}$ (BJD-2400000) & & &$54993.19529 \pm 0.00005$   \\
Mass ratio $q=M_2/M_1$               & & &$1.002 \pm 0.010$    \\ 
Orbital eccentricity, $e$              & & &$0.582\pm 0.002$     \\
Argument of periastron, $\omega_p$ (degree)              & & &$328.2\pm 0.7$     \\
$\gamma $ velocity (km s$^{-1}$)& & &$-0.81\pm 0.09$\\
Orbital inclination (degree), $i$ & & & $75.81\pm 0.34$\\
Semi-major axis ($R_\odot$), $a$ & & & $39.09\pm 0.16$\\
Mass ($M_\odot$)              & $2.05 \pm 0.03$             & $2.05 \pm 0.03$    \\ 
Radius ($R_\odot$)               & $2.96\pm 0.04$       & $2.51 \pm 0.05$    \\
Gravity brightening, $\beta$     &$0.25\tablenotemark{a}$    &$0.25\tablenotemark{a}$\\

Bolometric albedo & $1.0\tablenotemark{a}$   & $1.0\tablenotemark{a}$    \\

$T_{\rm eff}$ (K)                    & $7327\tablenotemark{a} $ & $7383 \pm 67$\\
$\log g$ (cgs)  & $3.81\pm 0.01$     & $3.95\pm 0.01$     \\ 
Model $v \sin i$ (km s$^{-1}$)   &   $8.67 \pm 0.8$   &   $7.35 \pm 0.9$              \\ 
Velocity semiamplitude $K$ (km~s$^{-1})$  &$84.4\pm 0.5 $ & $84.2\pm 0.7$\\ 
\enddata 
\tablenotetext{a}{Fixed.}
\end{deluxetable} 

\begin{deluxetable}{cccccccc}
\tabletypesize{\small} 
\tablewidth{0pc} 
\tablenum{3} 
\tablecaption{Radial Velocities\label{tab1}} 
\tablehead{ 
\colhead{Time}          & 
\colhead{Phase}&
\colhead{$V_{r}$(primary)}        & 
\colhead{$O-C$}        & 
\colhead{$V_{r}$(secondary)}    &
\colhead{$O-C$}        &                      \\  
\colhead{(BJD-2400000)}           & 
\colhead{}                 &
\colhead{(km s$^{-1}$)} & 
\colhead{(km s$^{-1}$)}& 
\colhead{(km s$^{-1}$)} & 
\colhead{(km s$^{-1}$)}& 
       
} 

\startdata		
       57204.11617  &  0.970      & $       58.76$ $\pm$         0.10 &  -0.17   &        -59.67  $\pm$        0.17&       $0.73$    \\
       57988.97196&     0.048   & $       108.23$ $\pm$        0.22 &  -1.3    &        -111.88  $\pm$        0.40&      $-0.97$    \\
       57933.94649&    0.117    & $       56.92$ $\pm$   0.09      &  0.92   &        -57.45  $\pm$        0.16&      $0.03$    \\
       57207.02947&   0.178     & $       25.82$ $\pm$  0.11      &  -1.09   &        -28.13   $\pm$        0.19&      $0.33$    \\
       57207.92302&  0.242     & $       7.39$ $\pm$  0.13      &  -0.27  &        -9.27  $\pm$        0.13&      $-0.02$    \\
       57179.97262&    0.245    & $       6.01$ $\pm$  0.13      & -0.92    &        -9.38  $\pm$        0.13&      $-0.84$    \\
       57994.01878 &   0.409     & $       -20.79$ $\pm$   0.29      &  0.36  &         19.55  $\pm$        0.43&      $0.08$    \\
       57218.06944&   0.967     & $       53.51$ $\pm$  0.10      &  0.29  &        -54.76  $\pm$        0.16&      $-0.05$    \\
       57254.87567&  0.597      & $       -38.12$ $\pm$  0.10      &   0.10   &        35.95  $\pm$        0.16&      $-0.56$    \\
       57202.11153 &   0.827     & $       -39.19$ $\pm$  0.10      &  -0.02    &        37.28  $\pm$        0.16&      $-0.17$    \\
       57202.88850&   0.883    & $       -25.50$ $\pm$  0.10      &    1.47  &        26.82  $\pm$        0.15&      $1.54$    \\
       57230.93288&    0.886    & $       -27.00$ $\pm$  0.10      &   -1.44   &        23.98 $\pm$        0.17&      $0.10$    \\
       57203.86463&    0.952    & $        28.53 $ $\pm$  0.10      &   0.22  &        -29.64  $\pm$        0.18&      $0.23$    \\
\enddata 
\end{deluxetable}


\begin{deluxetable}{lccccccc} 
\tabletypesize{\footnotesize} 
\tablewidth{0pc} 
\tablenum{4} 
\tablecaption{Orbital Harmonic Frequencies  \label{tab5}} 
\tablehead{ 
\colhead{}   & 
\colhead{Frequency (d$^{-1}$)}      &
\colhead{Amplitude (mag)}      &
\colhead{Phase (rad/$2\pi$)} & 
\colhead{S/N} &
\colhead{$N=f/f_{orb}$} &
%
}
\startdata 
 & Significant TEOs with S/N $> 10$  & $ $ & $ $ & $-$ & $-$ \\
\hline
$f_{ 19}$ & $ 0.6430676\pm  0.0000050$ & $0.000995\pm 0.000037$ & $0.0304\pm0.0172$ & $  46.4$ & $9$ \\
$f_{ 20}$ & $ 0.5716042\pm  0.0000052$ & $0.001129\pm 0.000043$ & $0.0816\pm0.0178$ & $  44.9$ & $8$ \\
$f_{ 23}$ & $ 1.2146299\pm  0.0000063$ & $0.000325\pm 0.000015$ & $0.5515\pm0.0216$ & $  36.9$ & $17$ \\
$f_{ 28}$ & $ 1.0002741\pm  0.0000078$ & $0.000332\pm 0.000019$ & $0.5757\pm0.0268$ & $  29.8$ & $14$ \\
$f_{ 32}$ & $ 0.9288526\pm  0.0000095$ & $0.000304\pm 0.000021$ & $0.5657\pm0.0324$ & $  24.6$ & $13$ \\
$f_{ 55}$ & $ 0.8573892\pm  0.0000129$ & $0.000252\pm 0.000024$ & $0.5796\pm0.0441$ & $  18.1$ & $12$ \\
$f_{ 71}$ & $ 0.7145309\pm  0.0000171$ & $0.000251\pm 0.000031$ & $0.0533\pm0.0584$ & $  13.7$ & $10$ \\
$f_{ 80}$ & $ 1.2861074\pm  0.0000185$ & $0.000105\pm 0.000014$ & $0.5394\pm0.0630$ & $  12.7$ & $18$ \\
$f_{ 82}$ & $ 1.4289999\pm  0.0000185$ & $0.000096\pm 0.000013$ & $0.5063\pm0.0633$ & $  12.6$ & $20$ \\
$f_{ 88}$ & $ 0.2143559\pm  0.0000195$ & $0.000525\pm 0.000075$ & $0.2248\pm0.0666$ & $  12.0$ & $3$ (artifact) \\
$f_{ 92}$ & $ 1.7148191\pm  0.0000202$ & $0.000078\pm 0.000012$ & $0.4800\pm0.0689$ & $  11.6$ & $24$ \\
\hline
\hline
 & Orbital Harmonic Peaks with S/N $< 10$  & $ $ & $ $ & $$ & $$ \\
\hline
$f_{104}$ & $ 1.0717173\pm  0.0000238$ & $0.000100\pm 0.000017$ & $0.5756\pm0.0812$ & $   9.8$ & $15$ \\
$f_{110}$ & $ 1.3575708\pm  0.0000254$ & $0.000073\pm 0.000014$ & $0.5112\pm0.0867$ & $   9.2$ & $19$ \\
$f_{112}$ & $ 1.6433901\pm  0.0000258$ & $0.000063\pm 0.000012$ & $0.4746\pm0.0881$ & $   9.1$ & $23$ \\
$f_{118}$ & $ 0.5001408\pm  0.0000282$ & $0.000243\pm 0.000050$ & $0.8610\pm0.0962$ & $   8.3$ & $7$ \\
$f_{119}$ & $ 1.1431464\pm  0.0000286$ & $0.000077\pm 0.000016$ & $0.5870\pm0.0976$ & $   8.2$ & $16$ \\
$f_{120}$ & $ 1.5004975\pm  0.0000288$ & $0.000059\pm 0.000013$ & $0.4680\pm0.0984$ & $   8.1$ & $21$ \\
$f_{125}$ & $ 1.5719430\pm  0.0000334$ & $0.000050\pm 0.000012$ & $0.4825\pm0.1141$ & $   7.0$ & $22$ \\
$f_{129}$ & $ 1.8577280\pm  0.0000405$ & $0.000038\pm 0.000011$ & $0.5515\pm0.1383$ & $   5.8$ & $26$ \\
$f_{130}$ & $ 1.7862988\pm  0.0000410$ & $0.000038\pm 0.000011$ & $0.4878\pm0.1400$ & $   5.7$ & $25$ \\
$f_{132}$ & $ 1.9291914\pm  0.0000418$ & $0.000036\pm 0.000011$ & $0.5941\pm0.1427$ & $   5.6$ & $27$ \\
$f_{134}$ & $ 0.0713949\pm  0.0000439$ & $0.000242\pm 0.000078$ & $0.3685\pm0.1498$ & $   5.3$ & $1$ \\
$f_{146}$ & $ 0.2858535\pm  0.0000579$ & $0.000167\pm 0.000071$ & $0.2820\pm0.1976$ & $   4.0$ & $4$ \\
\hline
$f_{      orb}$    & $ 0.071449999  \pm 0.0000003$ & $ -$ & $ -$ & $-$ & $-$ \\
\enddata 
\end{deluxetable}

 \begin{deluxetable}{lccccccc} 
\tabletypesize{\footnotesize} 
\tablewidth{0pc} 
\tablenum{5} 
\tablecaption{Non-orbital-harmonic Pulsations \label{tab5}} 
\tablehead{ 
\colhead{}   & 
\colhead{Frequency (day$^{-1}$)}      &
\colhead{Amplitude (mag)}      &
\colhead{Phase (rad/$2\pi$)} & 
\colhead{S/N} &
\colhead{Period (days)} &
%
}
\startdata 
$f_{  1}$ & $15.3552361\pm  0.0000006$ & $0.002211\pm 0.000010$ & $0.8816\pm0.0020$ & $ 392.4$ & $$ \\
$f_{  2}$ & $18.1528549\pm  0.0000008$ & $0.001627\pm 0.000010$ & $0.6811\pm0.0028$ & $ 288.8$ & $$ \\
$f_{  3}$ & $15.7633839\pm  0.0000010$ & $0.001311\pm 0.000010$ & $0.3349\pm0.0034$ & $ 232.6$ & $$ \\
$f_{  4}$ & $17.8741264\pm  0.0000012$ & $0.001057\pm 0.000010$ & $0.1297\pm0.0043$ & $ 187.6$ & $$ \\
$f_{  5}$ & $17.7825527\pm  0.0000013$ & $0.000980\pm 0.000010$ & $0.3382\pm0.0046$ & $ 174.0$ & $$ \\
$f_{  6}$ & $15.7641716\pm  0.0000014$ & $0.000927\pm 0.000010$ & $0.0244\pm0.0049$ & $ 164.5$ & $$ \\
$f_{  7}$ & $18.1515179\pm  0.0000017$ & $0.000782\pm 0.000010$ & $0.6120\pm0.0057$ & $ 138.8$ & $$ \\
$f_{  8}$ & $15.2402897\pm  0.0000018$ & $0.000740\pm 0.000010$ & $0.6525\pm0.0061$ & $ 131.3$ & $$ \\
$f_{  9}$ & $18.1404514\pm  0.0000022$ & $0.000599\pm 0.000010$ & $0.6132\pm0.0075$ & $ 106.3$ & $$ \\
$f_{ 10}$ & $18.1428509\pm  0.0000031$ & $0.000421\pm 0.000010$ & $0.2848\pm0.0107$ & $  74.8$ & $$ \\
$f_{ 11}$ & $17.7878971\pm  0.0000033$ & $0.000397\pm 0.000010$ & $0.2532\pm0.0113$ & $  70.4$ & $$ \\
$f_{ 12}$ & $17.8670006\pm  0.0000034$ & $0.000387\pm 0.000010$ & $0.5432\pm0.0116$ & $  68.8$ & $$ \\
$f_{ 13}$ & $18.1447697\pm  0.0000035$ & $0.000371\pm 0.000010$ & $0.4180\pm0.0121$ & $  65.9$ & $$ \\
$f_{ 14}$ & $15.2496424\pm  0.0000038$ & $0.000346\pm 0.000010$ & $0.4950\pm0.0130$ & $  61.4$ & $$ \\
$f_{ 15}$ & $18.1595688\pm  0.0000039$ & $0.000342\pm 0.000010$ & $0.4530\pm0.0132$ & $  60.7$ & $$ \\
$f_{ 16}$ & $18.1523056\pm  0.0000039$ & $0.000335\pm 0.000010$ & $0.1786\pm0.0134$ & $  59.4$ & $$ \\
$f_{ 17}$ & $15.2433386\pm  0.0000042$ & $0.000315\pm 0.000010$ & $0.5725\pm0.0143$ & $  55.9$ & $$ \\
$f_{ 18}$ & $17.1373272\pm  0.0000044$ & $0.000301\pm 0.000010$ & $0.6055\pm0.0149$ & $  53.4$ & $$ \\
$f_{ 21}$ & $17.8735104\pm  0.0000054$ & $0.000242\pm 0.000010$ & $0.4409\pm0.0186$ & $  42.9$ & $$ \\
$f_{ 22}$ & $ 0.6679736\pm  0.0000056$ & $0.000840\pm 0.000035$ & $0.2513\pm0.0193$ & $  41.4$ & $1.497065\pm 0.000013$ \\
$f_{ 24}$ & $18.1384659\pm  0.0000065$ & $0.000202\pm 0.000010$ & $0.6056\pm0.0223$ & $  35.8$ & $$ \\
$f_{ 25}$ & $17.8593616\pm  0.0000076$ & $0.000172\pm 0.000010$ & $0.2580\pm0.0261$ & $  30.6$ & $$ \\
$f_{ 26}$ & $17.8747082\pm  0.0000077$ & $0.000170\pm 0.000010$ & $0.8167\pm0.0264$ & $  30.2$ & $$ \\
$f_{ 27}$ & $15.7953129\pm  0.0000078$ & $0.000170\pm 0.000010$ & $0.8453\pm0.0265$ & $  30.2$ & $$ \\
$f_{ 29}$ & $18.1412067\pm  0.0000081$ & $0.000164\pm 0.000010$ & $0.6697\pm0.0275$ & $  29.0$ & $$ \\
$f_{ 30}$ & $18.1509705\pm  0.0000084$ & $0.000157\pm 0.000010$ & $0.1064\pm0.0286$ & $  27.9$ & $$ \\
$f_{ 31}$ & $15.3211060\pm  0.0000088$ & $0.000149\pm 0.000010$ & $0.3404\pm0.0302$ & $  26.4$ & $$ \\
$f_{ 33}$ & $ 0.6240198\pm  0.0000097$ & $0.000541\pm 0.000038$ & $0.1444\pm0.0330$ & $  24.2$ & $1.602513 \pm 0.000025$ \\
$f_{ 34}$ & $15.2411461\pm  0.0000097$ & $0.000136\pm 0.000010$ & $0.9669\pm0.0330$ & $  24.1$ & $$ \\
$f_{ 35}$ & $ 0.7005877\pm  0.0000098$ & $0.000450\pm 0.000032$ & $0.8063\pm0.0335$ & $  23.8$ & $1.427373\pm 0.000020$ \\
$f_{ 36}$ & $ 0.4454299\pm  0.0000100$ & $0.000768\pm 0.000056$ & $0.9656\pm0.0340$ & $  23.5$ & $2.245022\pm 0.000050$ \\
$f_{ 37}$ & $18.1602192\pm  0.0000100$ & $0.000132\pm 0.000010$ & $0.6743\pm0.0341$ & $  23.4$ & $$ \\
$f_{ 38}$ & $17.7868690\pm  0.0000102$ & $0.000129\pm 0.000010$ & $0.0400\pm0.0348$ & $  22.9$ & $$ \\
$f_{ 39}$ & $16.8639107\pm  0.0000104$ & $0.000126\pm 0.000010$ & $0.0115\pm0.0356$ & $  22.4$ & $$ \\
$f_{ 40}$ & $ 0.5118573\pm  0.0000104$ & $0.000641\pm 0.000049$ & $0.5020\pm0.0356$ & $  22.4$ & $1.953670\pm 0.000040$ \\
$f_{ 41}$ & $ 0.8546142\pm  0.0000107$ & $0.000306\pm 0.000024$ & $0.8962\pm0.0365$ & $  21.9$ & $1.170119\pm 0.000015$ \\
$f_{ 42}$ & $ 0.6835955\pm  0.0000107$ & $0.000426\pm 0.000034$ & $0.4573\pm0.0367$ & $  21.7$ & $1.462854\pm 0.000023$ \\
$f_{ 43}$ & $18.1457272\pm  0.0000108$ & $0.000122\pm 0.000010$ & $0.6132\pm0.0370$ & $  21.6$ & $$ \\
$f_{ 44}$ & $ 1.1305251\pm  0.0000108$ & $0.000206\pm 0.000016$ & $0.3245\pm0.0370$ & $  21.6$ & $0.884539 \pm 0.000008$ \\
$f_{ 45}$ & $17.8026962\pm  0.0000110$ & $0.000120\pm 0.000010$ & $0.6494\pm0.0375$ & $  21.3$ & $$ \\
$f_{ 46}$ & $15.7238493\pm  0.0000110$ & $0.000120\pm 0.000010$ & $0.3535\pm0.0375$ & $  21.3$ & $$ \\
$f_{ 47}$ & $15.3557749\pm  0.0000112$ & $0.000118\pm 0.000010$ & $0.4932\pm0.0381$ & $  20.9$ & $$ \\
$f_{ 48}$ & $ 0.9723609\pm  0.0000114$ & $0.000238\pm 0.000020$ & $0.6737\pm0.0389$ & $  20.5$ & $$ \\
$f_{ 49}$ & $15.7373466\pm  0.0000116$ & $0.000113\pm 0.000010$ & $0.7712\pm0.0397$ & $  20.1$ & $$ \\
$f_{ 50}$ & $ 0.8713667\pm  0.0000117$ & $0.000271\pm 0.000023$ & $0.3728\pm0.0400$ & $  19.9$ & $1.147623\pm 0.000015$ \\
$f_{ 51}$ & $ 0.4151111\pm  0.0000118$ & $0.000688\pm 0.000059$ & $0.1985\pm0.0401$ & $  19.9$ & $2.408994\pm 0.000068$ \\
$f_{ 52}$ & $17.8539829\pm  0.0000119$ & $0.000110\pm 0.000010$ & $0.1243\pm0.0408$ & $  19.6$ & $$ \\
$f_{ 53}$ & $15.3086348\pm  0.0000120$ & $0.000110\pm 0.000010$ & $0.8763\pm0.0410$ & $  19.5$ & $$ \\
$f_{ 54}$ & $15.3925686\pm  0.0000122$ & $0.000108\pm 0.000010$ & $0.8223\pm0.0416$ & $  19.2$ & $$ \\
$f_{ 56}$ & $ 0.8214177\pm  0.0000134$ & $0.000258\pm 0.000025$ & $0.3256\pm0.0458$ & $  17.4$ & $$ \\
$f_{ 57}$ & $18.0688171\pm  0.0000136$ & $0.000097\pm 0.000010$ & $0.7835\pm0.0463$ & $  17.2$ & $$ \\
$f_{ 58}$ & $ 1.0341558\pm  0.0000138$ & $0.000181\pm 0.000018$ & $0.4734\pm0.0470$ & $  17.0$ & $$ \\
$f_{ 59}$ & $15.3014412\pm  0.0000141$ & $0.000094\pm 0.000010$ & $0.3541\pm0.0480$ & $  16.6$ & $$ \\
$f_{ 60}$ & $15.8088446\pm  0.0000146$ & $0.000090\pm 0.000010$ & $0.2469\pm0.0499$ & $  16.0$ & $$ \\
$f_{ 61}$ & $ 0.5027102\pm  0.0000147$ & $0.000462\pm 0.000050$ & $0.8484\pm0.0503$ & $  15.9$ & $ 1.989218\pm 0.000058$ \\
$f_{ 62}$ & $18.1502495\pm  0.0000148$ & $0.000089\pm 0.000010$ & $0.0671\pm0.0505$ & $  15.8$ & $$ \\
$f_{ 63}$ & $18.1442738\pm  0.0000156$ & $0.000084\pm 0.000010$ & $0.9591\pm0.0533$ & $  15.0$ & $$ \\
$f_{ 64}$ & $17.8580246\pm  0.0000156$ & $0.000084\pm 0.000010$ & $0.3630\pm0.0534$ & $  14.9$ & $$ \\
$f_{ 65}$ & $18.1418571\pm  0.0000162$ & $0.000081\pm 0.000010$ & $0.0482\pm0.0553$ & $  14.4$ & $$ \\
$f_{ 66}$ & $15.3033943\pm  0.0000163$ & $0.000081\pm 0.000010$ & $0.0814\pm0.0555$ & $  14.4$ & $$ \\
$f_{ 67}$ & $18.2119598\pm  0.0000166$ & $0.000079\pm 0.000010$ & $0.0643\pm0.0567$ & $  14.1$ & $$ \\
$f_{ 68}$ & $18.1435184\pm  0.0000166$ & $0.000079\pm 0.000010$ & $0.7024\pm0.0568$ & $  14.0$ & $$ \\
$f_{ 69}$ & $17.1448479\pm  0.0000167$ & $0.000079\pm 0.000010$ & $0.4418\pm0.0571$ & $  14.0$ & $$ \\
$f_{ 70}$ & $17.7802753\pm  0.0000168$ & $0.000078\pm 0.000010$ & $0.9981\pm0.0574$ & $  13.9$ & $$ \\
$f_{ 72}$ & $ 0.2886627\pm  0.0000172$ & $0.000561\pm 0.000071$ & $0.7201\pm0.0588$ & $  13.6$ & $$ \\
$f_{ 73}$ & $ 1.0268604\pm  0.0000177$ & $0.000142\pm 0.000018$ & $0.7097\pm0.0605$ & $  13.2$ & $$ \\
$f_{ 74}$ & $15.7647543\pm  0.0000177$ & $0.000074\pm 0.000010$ & $0.7023\pm0.0606$ & $  13.2$ & $$ \\
$f_{ 75}$ & $ 1.1042287\pm  0.0000179$ & $0.000128\pm 0.000017$ & $0.7369\pm0.0611$ & $  13.1$ & $0.905587\pm 0.000015$ \\
$f_{ 76}$ & $15.2447767\pm  0.0000180$ & $0.000073\pm 0.000010$ & $0.2145\pm0.0615$ & $  13.0$ & $$ \\
$f_{ 77}$ & $15.8759918\pm  0.0000183$ & $0.000072\pm 0.000010$ & $0.2449\pm0.0623$ & $  12.8$ & $$ \\
$f_{ 78}$ & $15.2951384\pm  0.0000183$ & $0.000072\pm 0.000010$ & $0.0434\pm0.0626$ & $  12.8$ & $$ \\
$f_{ 79}$ & $ 1.1064554\pm  0.0000184$ & $0.000124\pm 0.000017$ & $0.5672\pm0.0630$ & $  12.7$ & $$ \\
$f_{ 81}$ & $ 0.1798575\pm  0.0000185$ & $0.000564\pm 0.000076$ & $0.9948\pm0.0630$ & $  12.7$ & $$ \\
$f_{ 83}$ & $17.9969959\pm  0.0000188$ & $0.000070\pm 0.000010$ & $0.0745\pm0.0643$ & $  12.4$ & $$ \\
$f_{ 84}$ & $ 0.3335414\pm  0.0000191$ & $0.000481\pm 0.000067$ & $0.7356\pm0.0651$ & $  12.3$ & $$ \\
$f_{ 85}$ & $15.3057585\pm  0.0000193$ & $0.000068\pm 0.000010$ & $0.6022\pm0.0658$ & $  12.1$ & $$ \\
$f_{ 86}$ & $15.2466621\pm  0.0000193$ & $0.000068\pm 0.000010$ & $0.1282\pm0.0660$ & $  12.1$ & $$ \\
$f_{ 87}$ & $15.8348475\pm  0.0000194$ & $0.000068\pm 0.000010$ & $0.7580\pm0.0663$ & $  12.0$ & $$ \\
$f_{ 89}$ & $15.8356686\pm  0.0000197$ & $0.000067\pm 0.000010$ & $0.4560\pm0.0672$ & $  11.9$ & $$ \\
$f_{ 90}$ & $17.7831192\pm  0.0000200$ & $0.000066\pm 0.000010$ & $0.6523\pm0.0684$ & $  11.7$ & $$ \\
$f_{ 91}$ & $15.6927423\pm  0.0000201$ & $0.000066\pm 0.000010$ & $0.0772\pm0.0685$ & $  11.6$ & $$ \\
$f_{ 93}$ & $ 0.5857188\pm  0.0000204$ & $0.000279\pm 0.000042$ & $0.0764\pm0.0695$ & $  11.5$ & $1.707404\pm 0.000059$ \\
$f_{ 94}$ & $17.1346035\pm  0.0000205$ & $0.000064\pm 0.000010$ & $0.8441\pm0.0700$ & $  11.4$ & $$ \\
$f_{ 95}$ & $15.2330608\pm  0.0000206$ & $0.000064\pm 0.000010$ & $0.2491\pm0.0703$ & $  11.3$ & $$ \\
$f_{ 96}$ & $18.0813732\pm  0.0000207$ & $0.000064\pm 0.000010$ & $0.9660\pm0.0707$ & $  11.3$ & $$ \\
$f_{ 97}$ & $17.1303558\pm  0.0000214$ & $0.000062\pm 0.000010$ & $0.7106\pm0.0731$ & $  10.9$ & $$ \\
$f_{ 98}$ & $15.8841791\pm  0.0000216$ & $0.000061\pm 0.000010$ & $0.3616\pm0.0738$ & $  10.8$ & $$ \\
$f_{ 99}$ & $15.2123003\pm  0.0000219$ & $0.000060\pm 0.000010$ & $0.8642\pm0.0748$ & $  10.7$ & $$ \\
$f_{100}$ & $15.2353220\pm  0.0000221$ & $0.000060\pm 0.000010$ & $0.2183\pm0.0754$ & $  10.6$ & $$ \\
$f_{101}$ & $18.1535568\pm  0.0000226$ & $0.000058\pm 0.000010$ & $0.4634\pm0.0772$ & $  10.3$ & $$ \\
$f_{102}$ & $15.6920233\pm  0.0000232$ & $0.000057\pm 0.000010$ & $0.3470\pm0.0793$ & $  10.1$ & $$ \\
$f_{103}$ & $15.3833189\pm  0.0000234$ & $0.000056\pm 0.000010$ & $0.3437\pm0.0799$ & $  10.0$ & $$ \\
$f_{105}$ & $15.4266901\pm  0.0000242$ & $0.000054\pm 0.000010$ & $0.2899\pm0.0826$ & $   9.7$ & $$ \\
$f_{106}$ & $15.7798281\pm  0.0000245$ & $0.000054\pm 0.000010$ & $0.9866\pm0.0837$ & $   9.5$ & $$ \\
$f_{107}$ & $17.7955551\pm  0.0000246$ & $0.000053\pm 0.000010$ & $0.4044\pm0.0841$ & $   9.5$ & $$ \\
$f_{108}$ & $11.0047159\pm  0.0000248$ & $0.000053\pm 0.000010$ & $0.8844\pm0.0846$ & $   9.4$ & $$ \\
$f_{109}$ & $17.1315212\pm  0.0000249$ & $0.000053\pm 0.000010$ & $0.7329\pm0.0852$ & $   9.4$ & $$ \\
$f_{111}$ & $ 1.3430794\pm  0.0000256$ & $0.000073\pm 0.000014$ & $0.8243\pm0.0875$ & $   9.1$ & $$ \\
$f_{113}$ & $15.4547138\pm  0.0000262$ & $0.000050\pm 0.000010$ & $0.3709\pm0.0896$ & $   8.9$ & $$ \\
$f_{114}$ & $15.7627668\pm  0.0000262$ & $0.000050\pm 0.000010$ & $0.6837\pm0.0896$ & $   8.9$ & $$ \\
$f_{115}$ & $ 0.6542016\pm  0.0000277$ & $0.000176\pm 0.000036$ & $0.6364\pm0.0946$ & $   8.4$ & $1.528661\pm 0.000065$ \\
$f_{116}$ & $ 0.7863309\pm  0.0000278$ & $0.000133\pm 0.000027$ & $0.5701\pm0.0949$ & $   8.4$ & $$ \\
$f_{117}$ & $ 0.9728688\pm  0.0000279$ & $0.000097\pm 0.000020$ & $0.0685\pm0.0953$ & $   8.4$ & $$ \\
$f_{121}$ & $ 0.3068540\pm  0.0000298$ & $0.000318\pm 0.000069$ & $0.2737\pm0.1017$ & $   7.8$ & $$ \\
$f_{122}$ & $ 0.2717732\pm  0.0000304$ & $0.000323\pm 0.000072$ & $0.6930\pm0.1037$ & $   7.7$ & $3.67954\pm 0.00041$ \\
$f_{123}$ & $ 1.1549314\pm  0.0000313$ & $0.000069\pm 0.000016$ & $0.4031\pm0.1069$ & $   7.5$ & $0.865780\pm 0.000023$ \\
$f_{124}$ & $ 1.0336092\pm  0.0000331$ & $0.000075\pm 0.000018$ & $0.0818\pm0.1129$ & $   7.1$ & $$ \\
$f_{126}$ & $ 0.3915069\pm  0.0000337$ & $0.000250\pm 0.000062$ & $0.5218\pm0.1150$ & $   6.9$ & $$ \\
$f_{127}$ & $ 0.7391911\pm  0.0000343$ & $0.000119\pm 0.000030$ & $0.5125\pm0.1170$ & $   6.8$ & $ 1.352882\pm  0.000063$ \\
$f_{128}$ & $ 0.4404624\pm  0.0000401$ & $0.000192\pm 0.000057$ & $0.3691\pm0.1370$ & $   5.8$ & $2.27016\pm 0.00021$ \\
$f_{131}$ & $ 0.6038013\pm  0.0000411$ & $0.000133\pm 0.000040$ & $0.8542\pm0.1402$ & $   5.7$ & $$ \\
$f_{133}$ & $ 1.3077447\pm  0.0000427$ & $0.000045\pm 0.000014$ & $0.5128\pm0.1458$ & $   5.5$ & $$ \\
$f_{135}$ & $ 0.7381976\pm  0.0000454$ & $0.000090\pm 0.000030$ & $0.2275\pm0.1551$ & $   5.1$ & $$ \\
$f_{136}$ & $ 0.8451529\pm  0.0000477$ & $0.000070\pm 0.000024$ & $0.4170\pm0.1629$ & $   4.9$ & $$ \\
$f_{137}$ & $ 0.9362464\pm  0.0000492$ & $0.000058\pm 0.000021$ & $0.8067\pm0.1679$ & $   4.8$ & $$ \\
$f_{138}$ & $ 0.3329590\pm  0.0000502$ & $0.000183\pm 0.000067$ & $0.3079\pm0.1714$ & $   4.7$ & $$ \\
$f_{139}$ & $ 1.1291689\pm  0.0000507$ & $0.000044\pm 0.000016$ & $0.5154\pm0.1730$ & $   4.6$ & $$ \\
$f_{140}$ & $ 0.8705727\pm  0.0000507$ & $0.000063\pm 0.000023$ & $0.4639\pm0.1731$ & $   4.6$ & $$ \\
$f_{141}$ & $ 0.2691011\pm  0.0000524$ & $0.000188\pm 0.000072$ & $0.8015\pm0.1789$ & $   4.5$ & $$ \\
$f_{142}$ & $ 1.1282299\pm  0.0000524$ & $0.000043\pm 0.000016$ & $0.6983\pm0.1790$ & $   4.5$ & $$ \\
$f_{143}$ & $ 1.8098345\pm  0.0000533$ & $0.000029\pm 0.000011$ & $0.3362\pm0.1821$ & $   4.4$ & $$ \\
$f_{144}$ & $ 0.4380926\pm  0.0000535$ & $0.000145\pm 0.000057$ & $0.6083\pm0.1826$ & $   4.4$ & $$ \\
$f_{145}$ & $ 1.5049332\pm  0.0000571$ & $0.000030\pm 0.000012$ & $0.2163\pm0.1949$ & $   4.1$ & $$ \\
$f_{147}$ & $ 0.4030521\pm  0.0000583$ & $0.000142\pm 0.000060$ & $0.7683\pm0.1992$ & $   4.0$ & $$ \\
$f_{148}$ & $ 0.2284019\pm  0.0000584$ & $0.000174\pm 0.000074$ & $0.4863\pm0.1992$ & $   4.0$ & $$ \\
\enddata 
\end{deluxetable}

\end{document}